\documentclass[11pt,a4paper,nofootinbib,superscriptaddress]{revtex4-1}

\pdfoutput=1
\usepackage[utf8]{inputenc}
\usepackage{graphicx,bm,bbm}
\usepackage{slashed,verbatim}
\usepackage{amssymb,graphicx,epstopdf}
\usepackage{slashed,subfigure}
\usepackage{caption}
\usepackage{amsmath,mathtools}
\usepackage{epstopdf,dcolumn}
\usepackage{soul}

\allowdisplaybreaks
\usepackage[normalem]{ulem}
\usepackage{cancel}
\usepackage{xcolor}
 \usepackage[colorlinks=true,linktocpage=true,linkcolor=blue,citecolor=blue]{hyperref}
\usepackage{xcolor}
\allowdisplaybreaks

\def\be {\begin{equation}}
\def\ee {\end{equation}}
\def\bea {\begin{eqnarray}}
\def\eea {\end{eqnarray}}
\def\bc {\begin{center}}
\def\ec {\end{center}}
\def\nn {\nonumber}
\def\eps {\epsilon}
\def\gm {\gamma}

\def\mn {\mu\nu}
\def\({\left(}
\def\){\right)}
\def\[{\left[}
\def\]{\right]}
\def\sp {\shortparallel}
\newcommand \Tr{\operatorname{\text{Tr}}}

\def\sumintb{\sum\!\!\!\!\!\!\!\!\!\int\limits}
\def\sumintf{\sum\!\!\!\!\!\!\!\!\!\!\int\limits}

\def\slashed{\slash\!\!\!\!}
\def\slashedl{\slash\!\!\!}

\DeclareGraphicsExtensions{.jpg,.pdf,.eps}

\begin{document}
\title{Anisotropic pressure of deconfined QCD matter in presence of strong magnetic field within one-loop approximation}
\author{Bithika Karmakar}
\email{bithika.karmakar@saha.ac.in}
  \affiliation{
 	Theory Division, Saha Institute of Nuclear Physics, HBNI, \\
 	1/AF, Bidhannagar, Kolkata 700064, India}
 	\author{Ritesh Ghosh}
\email{ritesh.ghosh@saha.ac.in}
  \affiliation{
 	Theory Division, Saha Institute of Nuclear Physics, HBNI, \\
 	1/AF, Bidhannagar, Kolkata 700064, India}
 \author{Aritra Bandyopadhyay}
  \email{aritrabanerjee.444@gmail.com}
 \affiliation{Departamento de F\'{\i}sica, Universidade Federal de Santa Maria, 
 	Santa Maria, RS 97105-900, Brazil}
	 
 \author{Najmul Haque}
 \email{nhaque@niser.ac.in}
 \affiliation{School of Physical Sciences, National Institute of Science Education and Research, HBNI,\\  Jatni, Khurda 752050, India}
 
 \author{Munshi G Mustafa}
 \email{munshigolam.mustafa@saha.ac.in}
  \affiliation{
 	Theory Division, Saha Institute of Nuclear Physics, HBNI, \\
 	1/AF, Bidhannagar, Kolkata 700064, India}

\begin{abstract}{
 Considering the general structure of the two point functions of quarks and gluons, we compute the free energy and pressure of a strongly magnetized hot and dense QCD matter created in heavy-ion collisions. In the presence of a strong magnetic field we found that the deconfined QCD matter exhibits a paramagnetic nature. One gets different pressures in directions parallel and perpendicular to the magnetic field due to the magnetization acquired by the system. We obtain both longitudinal and transverse pressures, and magnetization of hot deconfined QCD matter in the presence of the magnetic field. We have used hard thermal loop approximation for the heat bath. We obtained completely analytic expressions for pressure and magnetization under certain approximations. Various divergences appearing in free energy are regulated using appropriate counterterms. The obtained anisotropic pressure  may be useful for a magnetohydrodynamics description of a hot and dense deconfined QCD matter produced in heavy-ion collisions.}

\end{abstract}

\maketitle 
\newpage
\section{Introduction}
Quantum chromodynamics (QCD) is the theory of the strong interaction that has two important features. One is the feeble interaction of quarks and gluons at high energy, and the other one is the confinement in which the interaction strength becomes strong at low energy. A transition between these two phases, namely confined to a deconfined state of hadronic matter known as quark-gluon plasma (QGP), is supposed to occur at around the energy scale or temperature $160$ MeV. In the early universe such a state of matter is presumed to be created after a few microseconds of big-bang. It can also exist in the core of neutron stars as matter density is much higher than the normal nuclear matter density. Various high energy heavy-ion experiments are underway in laboratories (LHC at CERN and RHIC at BNL) to study the formation of QGP and its properties for unraveling the characteristics of the  QCD phase diagram. Future experiments are also  planned in FAIR at GSI and NICA at Dubna to explore the high baryon density domain of the QCD phase diagram.

In recent years much attention has been paid in non-central heavy-ion collisions (HIC) where a magnetic field as high as $(10-30)m_{\pi}^2$ can be generated in a direction perpendicular to the reaction plane. This magnetic field is primarily created when the spectators recede from each other. This magnetic field strength, however, also decreases very fast to $(1-2) m_{\pi}^2$ after a timescale of $(4-5)$fm/c. The presence of an external magnetic field  introduces  an extra energy scale in the system in addition to the scales ($gT$ and $T$; $g$ is the strong coupling) associated with a heat bath. One can work with two limiting cases: 
the strong magnetic field limit $(eB>T^2)$ and the weak magnetic field limit $(eB< T^2)$. We note that the presence of an anisotropic magnetic field in the medium  requires an appropriate modification of the present theoretical tools to investigate various properties of QGP. In recent years numerous activities have been in progress such as  magnetic catalysis~\cite{Alexandre:2000yf,Gusynin:1997kj,Lee:1997zj}, inverse magnetic catalysis 
\cite{Bali:2011qj,Bornyakov:2013eya,Mueller:2015fka,Ayala:2014iba,Farias:2014eca,Ayala:2014gwa,Ayala:2016sln,Ayala:2015bgv,Mukherjee:2018ebw}
and chiral magnetic effect~\cite{Kharzeev:2007jp,Fukushima:2008xe,Kharzeev:2009fn} at finite temperature, and
the chiral- and color-symmetry broken/restoration 
phase~\cite{Avancini:2016fgq,Fayazbakhsh:2010bh,Fayazbakhsh:2010gc,Andersen:2012zc,Andersen:2014xxa}.
Also in progress is the study related to the equation of state (EoS) in thermal perturbative QCD (pQCD) models~\cite{Bandyopadhyay:2017cle,Rath:2017fdv},
holographic models~\cite{Rougemont:2015oea,Finazzo:2016mhm}  and various 
thermodynamic  properties~\cite{Andersen:2012zc,Andersen:2014xxa,Strickland:2012vu,Farias:2016gmy},
 refractive indices and decay 
constant of hadrons \cite{Fayazbakhsh:2012vr,Fayazbakhsh:2013cha, Bandyopadhyay:2016cpf,Bandyopadhyay:2018gbw,Chakraborty:2017vvg,Islam:2018sog,Ghosh:2018xhh,Ghosh:2019fet}; 
soft photon production from conformal anomaly~\cite{Basar:2012bp,Ayala:2016lvs} in 
HIC; modification of dispersion properties in a magnetized hot QED~\cite{Sadooghi:2015hha,Karmakar:2018aig} 
and QCD~\cite{Karmakar:2018aig, Ayala:2018ina,Das:2017vfh,Hattori:2017xoo} medium; and
various transport coefficients~\cite{Kurian:2018qwb,Kurian:2018dbn,Kurian:2017yxj},
properties of quarkonia~\cite{Singh:2017nfa,Hasan:2018kvx},
synchroton radiation~\cite{Tuchin:2013prc2}, dilepton 
production from a hot magnetized QCD 
plasma~\cite{Bandyopadhyay:2016fyd,Tuchin:2013prc,Tuchin:2013prc2,Tuchin:2013ie,
Sadooghi:2016jyf,Bandyopadhyay:2017raf} and in strongly coupled plasma in a strong magnetic 
field~\cite{Mamo:2013efa}.

The EoS is an important quantity and is of phenomenological importance for studying the hot and dense QCD matter,  QGP, created in the relativistic heavy-ion collisions. This is because the EoS determines the thermodynamic properties of a hot and dense medium. Also the  time evolution  of the hot and dense  fireball  is studied through hydrodynamic models that require an EoS of the deconfined QCD matter  as an input. In the absence of a magnetic field  the EoS has systematically been computed in lattice QCD (LQCD) \cite{Gunther:2016vcp,Bazavov:2017dus,Bazavov:2017dsy} and  in hard thermal loop perturbation theory (HTLpt) within two loop [next-to-leading order (NLO)]~\cite{Haque:2012my}  and
three loop [next-to-NLO (NNLO)]~\cite{3loopglue1,3loopglue2,3loopqcd1,3loopqcd2,3loopqcd3,najmul3,Haque:2014rua}  
at finite temperature and chemical potential. On the other hand as high magnetic fields are being produced in noncentral HIC, they subsequently decrease with the expansion of the fireball. Such systems, i.e. the expansion dynamics of a thermomagnetic medium, are governed by magnetohydrodynamics~\cite{Inghirami:2016iru,Roy:2017yvg} that require a magnetic field dependent EoS as an input. In view of this, a systematic  determination of EOS for a magnetized hot QCD medium is of great importance.  Some LQCD effort has been made in Ref.~\cite{Bali:2014kia} 
that is limited to the temperature range (100-300)MeV. Recently we have computed~\cite{Bandyopadhyay:2017cle} the thermomagnetic EoS for the hot magnetized QCD medium within
the weak magnetic field  and HTL approximation.  Also using HTL approximation, some  thermodynamic quantities in lowest Landau level (LLL) within the strong field approximation has been numerically computed in Ref.~\cite{Rath:2017fdv}.  However, for the gluonic case, this calculation assumes  that the only effect of the magnetic field is to shift the Debye mass without any change in the general structure of two point functions at the finite temperature. It has explicitly been shown~\cite{Karmakar:2018aig, Ayala:2018ina,Das:2017vfh,Hattori:2017xoo} that the presence of an external magnetic 
field breaks the rotational symmetry and the situation is quite different from what has been assumed in Ref.~\cite{Rath:2017fdv}. This seeks an improvement
of the general structure of two point functions used in Ref.~\cite{Rath:2017fdv}. 

In view of this we systematically compute
the EoS  within strong field approximation by exploiting the general structure of effective propagator of quarks and gluons in a thermomagnetic QCD medium. 
In the strong field limit, the magnetic 
field pushes higher  Landau levels (HLL)  to infinity compared to the lowest
 Landau level (LLL)~\cite{Bandyopadhyay:2016fyd}. Thus we work with the LLL approximation along with a scale hierarchy $eB>T^2>m_f^2$, where $m_f$ is the 
 intrinsic mass scale associated with quarks.  We also compute the magnetization  which indicates
that the deconfined hot and dense QCD matter is paramagnetic in nature. We further note that, in the presence of a strong magnetic field, we take into account the anisotropy between longitudinal and transverse pressures which is created due to the fact 
that  the system acquires a magnetization along the field direction and  is likely to elongate more along the direction of the magnetic field.   
We obtain a completely analytic expression for anisotropic (both longitudinal and transverse) pressures and magnetization under a certain approximation. 

The paper has been organized as follows. In Sec.~\ref{setup} we describe the basic setup for the computation of the free energy in this manuscript. In Sec.~\ref{fermion_gs}
we discuss the general structure of fermion self-energy in the presence of a strong magnetic field, the effective fermion propagator and associated  form factors, and the quark free energy
in one loop up to  $\mathcal O(g^4)$. The hard and soft contributions of gluon free energy up to $\mathcal O(g^4)$ are also calculated in Sec.~\ref{GF} within one-loop HTL approximation. In Sec.~\ref{thermo} the pressure of an anisotropic system is discussed. We discuss our results in  Sec.~\ref{res}. Finally, we conclude in Sec.~\ref{conclu}.  

\section{Setup}
\label{setup}
The total thermodynamic free energy up to one-loop order in HTLpt in the presence of a background magnetic field, $B$, can be written as
\bea
F &=& F_q+F_g+F_0 + \Delta {\mathcal E}^0_T +\Delta {\mathcal E}_T^B
\label{total_fe}
\eea
where $F_q$ and $F_g$ are, respectively, the quark and gluon part of the free energy which will be computed in presence of 
magnetic field with an HTL approximation. 
$F_0$ is the tree level contribution due to the constant magnetic field, given as 
\be
F_0\rightarrow\frac{1}{2} B^2 + \Delta {\mathcal E}_0^{B^2},
\ee
where $ \Delta {\mathcal E}_0^{B^2}$ is a counterterm of ${\mathcal O} {[(q_fB)^2]}$ from vacuum as we will see later.
The $\Delta {\mathcal E}_T$  is a counterterm independent of the magnetic field  (viz. ${\mathcal O}[(q_fB)^0 T^4]$ )as
\bea
\Delta {\mathcal E}^{0}_T&=& \Delta {\mathcal E}^{\mbox{\tiny{HTL}}}_T+\Delta {\mathcal E}_T,
\label{htl_count}
\eea
where $ \Delta {\mathcal E}^{\mbox{\tiny{HTL}}}_T$ is the HTL counterterm~\cite{najmul3,Haque:2014rua}.
The counterterm $\Delta {\mathcal E}_T$ arises due to  the quark loop in the gluonic two point function
in presence of magnetic field but the field dependence explicitly gets canceled from the denominator and numerator
as we will see later. Finally, the counterterm
$\Delta {\mathcal E}_T^B$  is of order ${\mathcal O}[(q_fB) T^2]$ and ${\mathcal O}[(q_fB)^3/ T^2]$ .

The pressure of a system is defined as 
\be
P=-F. \label{pressure}
\ee
We also note the QCD Casimir numbers are $C_A=N_c$, $d_A=N_c^2-1$, $d_F=N_cN_f$ and $C_F=(N_c^2-1)/2N_c$ where $N_c$ is the number of color
and $N_f$ is the number of quark flavor.

\section{Quarks in a strong magnetic field}
\label{fermion_gs}
\subsection{General structure of fermion self-energy in strong field approximation}

It is established that the presence of a heat bath breaks the Lorentz(boost) invariance, whereas the presence of a magnetic field, $B$,
breaks the rotational symmetry of the system. In such a situation one needs to construct a manifestly covariant structure of  the self-energy. 
The presence of a heat  bath introduces a four vector $u^\mu$,  which is the velocity of the heat bath in addition to $P^\mu$, 
the four momentum of the external fermion.  Now, in the case of noncentral heavy ion collisions, when QGP can be generally identified by the presence of external anisotropic magnetic and electric fields, the said heat bath is further considered to be in the vicinity of an external electromagnetic field. In this case one can construct two more four vectors in the comoving frame of the heat bath, i.e., $n^\mu$ and $e^\mu$, thereby characterizing the previously mentioned magnetic and electric fields, respectively. This is done by combining the electromagnetic field tensor $F^{\mu\nu}$  or its dual 
$\tilde{F}^{\mu\nu}$ with the fluid velocity $u^\mu$  as
\bea
 n_\mu \equiv \frac{1}{2B} \epsilon_{\mu\nu\rho\lambda}\, u^\nu F^{\rho\lambda} 
 &=& \frac{1}{B}u^\nu {\tilde F}_{\mu\nu} = (0,0,0,1), \label{bmu} \\
 e_\mu &=& \frac{1}{E}u^\nu  F_{\mu\nu} = (0,1,0,0),
 \eea
 where the external magnetic and electric fields are considered to be in the $z$ and $x$ directions, respectively.  This can further be justified by the structure of the electromagnetic field tensor
\bea
   F^{\mu\nu}=
  \left( {\begin{array}{cccc}
   0 & E & 0 & 0 \\
   -E & 0 & -B & 0 \\
   0 & B & 0 & 0 \\
    0 & 0 & 0 & 0 \\
  \end{array} } \right)
\eea
which has been projected out along the four velocity, i.e. the rest frame of heat bath  as  $u^\mu=(1,0,0,0)$. This also uniquely establishes a connection between the heat bath, the external magnetic field along the $z$ direction and the external electric field along the $x$ direction. The general expression of $F^{\mn}$ in terms of the four vectors $u^{\mu}$, $e^{\mu}$ and $n^{\mu}$ in the local rest frame of the heat bath is given by
\bea
F^{\mn}&=& E\big(e^{\mu} u^{\nu}-e^{\nu} u^{\mu}\big)+B \eps^{\mu\nu\rho\lambda}u_{\rho} n_{\lambda}.
\eea
At this point we would also like to mention that in the rest part of our manuscript, following the ideal magnetohydrodynamics approximation we consider $e_\mu = 0$, assuming the plasma in scrutiny has an infinite electrical conductivity~\cite{Inghirami:2016iru}.

The fermion self-energy $\Sigma(P)$ is  $4\times 4$ matrix that can be constructed from Dirac $\gamma$ matrices\footnotemark\footnotetext{
We note that the $\sigma^{\mu\nu}$ do not appear due to antisymmetric nature of it in any loop order of self-energy.}. The self-energy will also depend on four vectors $P^\mu$, $u^\mu$ and $n^{\mu}$. In the presence of the magnetic field one can generally define
\bea
P_\perp^\mu&=&P^\mu-(P\cdot u)u^\mu +(P\cdot n) n^\mu ,\\
P_\sp^\mu&=&(P\cdot u)u^\mu -(P\cdot n) n^\mu ,\\
p_\perp^2&=& (P^\mu u_\mu)^2-(P^\mu n_\mu)^2- P^\mu P_\mu \nn\\
&=&p_1^2+p_2^2 =-P_\perp^2.
\eea
As mentioned in the Introduction, in the present work we will be dealing with very strong external anisotropic magnetic fields, which are of relevance for initial stages of a noncentral heavy-ion collisions. In the presence of such strong magnetic fields ($q_fB \gg T^2$), we usually confine ourselves in the LLL. The reason for this assumption can be simply understood by looking at the dispersion relation in the presence of an external magnetic field along $z$ direction, i.e. $E_n=p_0 =\sqrt{p_3^2+m_f^2+2neB}$, with $n$ representing the number of Landau levels. As can be seen from the above expression, for LLL, i.e. for $n=0$, the dispersion relation is independent of $eB$; hence a higher value of $eB$ does not affect it, instead pushing all other HLLs ($n\neq 0$) far away from LLL (see Fig.1 of Ref~\cite{Bandyopadhyay:2016fyd}). So, for strong enough $eB$, the energy gap becomes too large for a fermion to hop out of LLL and realistically we can work only with LLL. In LLL with a strong field approximation ($q_fB\gg T^2 $), an effective dimensional reduction takes place 
for fermions from (3+1) to (1+1) whereas gluons move in usual (3+1) dimension~\cite{Gusynin:1995nb}. In addition  one has also a scale associated with the current quark mass $m_f$. In LLL  the electrical conductivity is very sensitive to $m_f$ as it becomes infinite in the massless limit. This is because in the massless limit ($m_f=0$) due to chirality conservation the scattering processes are forbidden such that the electrical conductivity diverges without scattering~\cite{Hattori:2016cnt}. We will also see below
in Eq.~\eqref {mass_debye} that the gluons acquire a screening mass 
$m_D^s \sim g\sqrt{q_fB}$ ($g$ is the QCD coupling). Since we are interested in finite  temperature effect, we can consider $T \gg m_D^s, m_f$. In this paper we will work
in the following hierarchy $q_fB \gg T^2 \gg (m_D^s)^2, m_f^2$.

In the LLL the transverse component of the momentum, $P_\perp=0$. Thus, $P^\mu$ reduces to $P_\sp^\mu$. $P_\sp^\mu$ can be written 
as a linear combination of $u^\mu$ and $n^\mu$. In the chiral limit the general structure of fermion self-energy in lowest Landau level can be written as
\bea
\Sigma(p_0,p_3)&=& a \slashedl u + b \slashedl n + c\gamma_5 \slashedl u +d\gamma_5 \slashedl n ,
\eea
where  $\slashedl{u}=\gamma_0$ and $\slashedl{n}=\gamma^3 n_3 =\gamma^3$. Now, the various form factors can be obtained as
\bea
a&=&\frac{1}{4}\Tr[\Sigma \slashedl u], \\
b &=& -\frac{1}{4}\Tr[\Sigma \slashedl n] , \\
c&= &\frac{1}{4}\Tr[ \gamma_5 \Sigma\slashedl u], \\
d&=&-\frac{1}{4}\Tr[ \gamma_5 \Sigma\slashedl n].
\eea
Finally, using the chirality projectors we can express the general structure of the fermion self-energy as,
\begin{align}
    \Sigma(P)=\mathcal{P}_R \ \slashed{A} \ \mathcal{P}_L+\mathcal{P}_L \ \slashed{B}\ \mathcal{P}_R  ,
\end{align} 
where 
\begin{align}
    &\slashed{A}=(a+c)\slashedl{u}+(b+d)\slashedl{n},\\
    &\slashed{B}=(a-c)\slashedl{u}+(b-d)\slashedl{n}, \\
&\mathcal{P}_R=\frac{1}{2}(\mathbbm{1}+\gamma_5) ,\\
&\mathcal{P}_L=\frac{1}{2}(\mathbbm{1}-\gamma_5).
\end{align}


\subsection{One loop quark self-energy in presence of a strong magnetic field}
\label{qse_htl_sfa} 
\begin{center}
\begin{figure}[tbh]
 \begin{center}
 \includegraphics[scale=0.4]{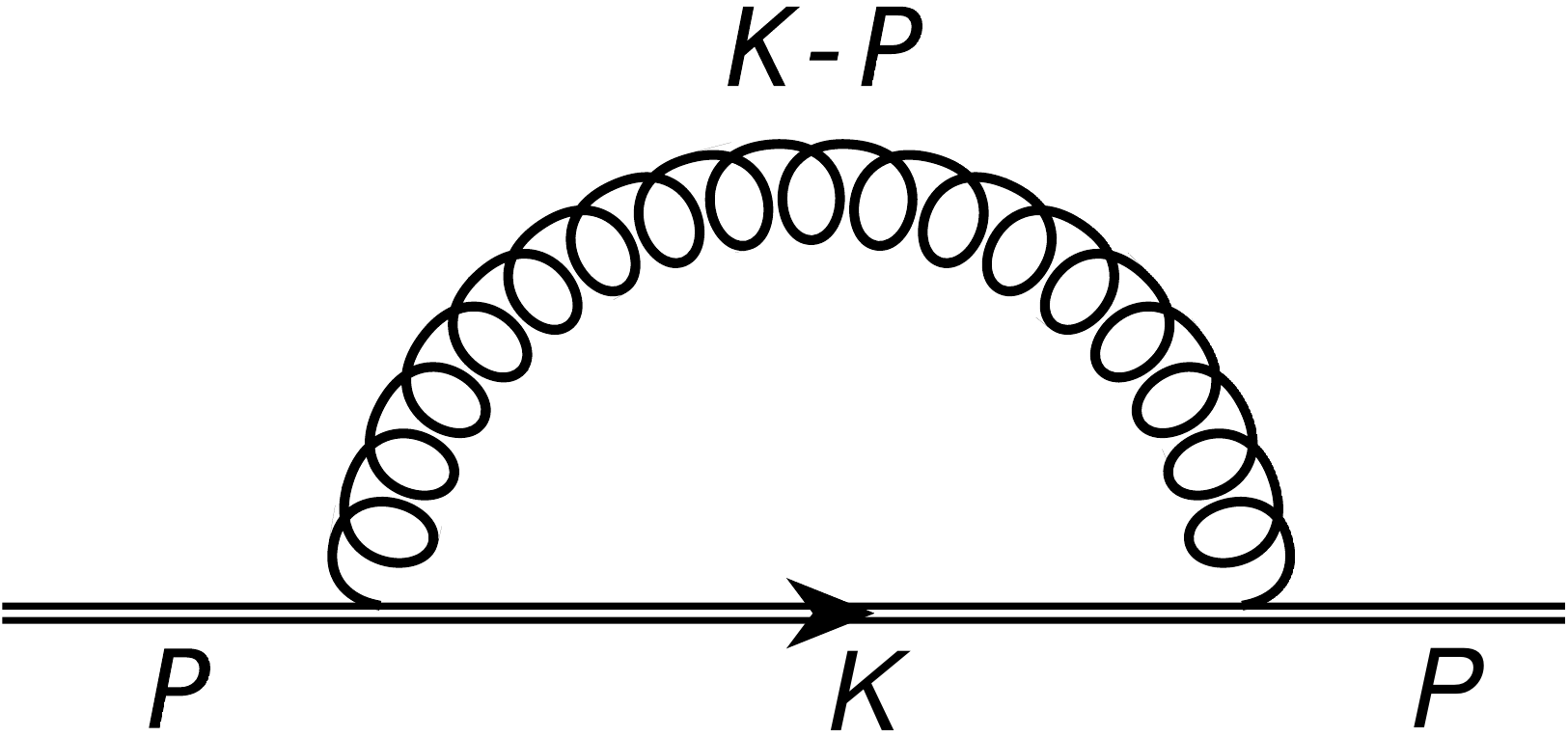} 
 \caption{Self-energy diagram for a quark in a strong magnetic field approximation. 
The double line indicates the modified quark propagator in presence of strong magnetic 
field.}
  \label{quark_se}
 \end{center}
\end{figure}
\end{center}

Using the modified fermion propagator in strong field approximation, one can right away write down the quark self-energy in Feynman gauge from Fig.~\ref{quark_se} as
\bea
\Sigma(P) &=& -ig^2C_F\int\frac{d^4K}{(2\pi)^4}\gamma_\mu S(K) \gamma^\mu 
\Delta(K-P),
\label{sfa_self}
\eea
where  the unmodified  gluonic propagator is given as
\bea
\Delta(K-P) = \frac{1}{(k_0-p_0)^2-(k-p)^2} =  \frac{1}{(K-P)_\sp^2-(k-p)_\perp^2}.
\eea  
and the modified fermion propagator in LLL is given by
\bea
iS(K)=ie^{-{k_\perp^2}/{q_fB}}~~\frac{\slashed{K}_\sp+m_f}{
K_\sp^2-m_f^2}\(\mathbbm{1}-i\gamma_1\gamma_2\).
\label{prop_sfa}
\eea

Now, the thermo-magnetic self-energy $\Sigma(P)$ can be written from 
Eq.~(\ref{sfa_self}) as
\bea
\Sigma(P) &=& -ig^2C_F\int\frac{d^4K}{(2\pi)^4} e^{-k_\perp^2/q_fB} \gamma_\mu\frac{(\slashed{K}_\sp 
+m_f)}{(K_\sp^2-m_f^2)}\(\mathbbm{1}-i\gm_1\gm_2\) \gamma^\mu \Delta(K-P) \nn\\
&=& -ig^2C_F\int\frac{d^4K}{(2\pi)^4}e^{-k_\perp^2/q_fB}\gamma_\mu \slashed{K}_\sp 
\(\mathbbm{1}-i\gm_1\gm_2\)\gamma^\mu 
\tilde{\Delta}_\sp(K)\Delta(K-P) \nn \\
&=& -2g^2C_F\sumintf_{\{ K\} }e^{-k_\perp^2/q_fB}\left[\(\mathbbm{1}+i\gm_1\gm_2\)\slashed{K}_\sp \right]\tilde{\Delta}_\sp(K)\Delta(K-P), \label{self_sfa_tot} 
\eea
where 
\bea
\tilde{\Delta}_\sp(K) = \frac{1}{k_0^2-k_3^2}  
\eea  
 Also at finite temperature, the loop integration measure is replaced by
\bea
\int\frac{d^4K}{(2\pi)^4}  \longrightarrow 
iT\sum\limits_{\{k_0\}}\int\frac{d^3k}{(2\pi)^3}\
\longrightarrow  iT\sum\limits_{\{k_0\}}\int\frac{dk_3}{2\pi}\int\frac{d^2k_\perp}{(2\pi)^2} \hspace{.2cm}.
\eea

Now the expressions of form factors for a particular flavor $f$ become
\bea
a=\frac{1}{4}\Tr[\Sigma\, \slashed{u}]&=&-\frac{2 g^2 C_F}{4}\sumintf_{\{ K\} } e^{-\frac{k_\perp^2}{q_fB}}\Tr\bigg[(\mathbbm{1}+i \gm_1\gm_2)\,\slashed{k}_\sp\,
\slashed{u}\bigg]\tilde{\Delta}_\sp(K)\Delta(K-P)\nn\\
&=&-2g^2 C_F \sumintf_{\{ K\} } e^{-\frac{k_\perp^2}{q_fB}}k^0\tilde{\Delta}_\sp(K)\Delta(K-P)\label{quark_a},
\eea

\bea
b=-\frac{1}{4}\Tr[\Sigma\: \slashed{n}]&=&\frac{2 g^2 C_F}{4}\sumintf_{\{ K\} } e^{-\frac{k_\perp^2}{q_fB}}\Tr\bigg[\(\mathbbm{1}+i \gm_1\gm_2\)\,\slashed{k}_\sp\,
\slashed{n}\bigg]\tilde{\Delta}_\sp(K)\Delta(K-P)\nn\\
&=&2g^2 C_F \sumintf_{\{ K\} } e^{-\frac{k_\perp^2}{q_fB}}k^3\tilde{\Delta}_\sp(K)\Delta(K-P)\label{quark_b},
\eea

\bea
c=\frac{1}{4}\Tr[\gm_5\Sigma\, \slashed{u}]&=&-\frac{2 g^2 C_F}{4}\sumintf_{\{ K\} } e^{-\frac{k_\perp^2}{q_fB}}\Tr\bigg[\gm_5 \(\mathbbm{1}+i \gm_1\gm_2\)\,\slashed{k}_\sp\,
\slashed{u}\bigg]\tilde{\Delta}_\sp(K)\Delta(K-P)\nn\\
&=&-2g^2 C_F \sumintf_{\{ K\}} e^{-\frac{k_\perp^2}{q_fB}}k^3\tilde{\Delta}_\sp(K)\Delta(K-P),
\eea

\bea
d=-\frac{1}{4}\Tr[\gm_5\Sigma\, \slashed{n}]&=&\frac{2 g^2 C_F}{4}\sumintf_{\{ K\} } e^{-\frac{k_\perp^2}{q_fB}}\Tr\bigg[\gm_5(\mathbbm{1}+i \gm_1\gm_2)\,\slashed{k}_\sp\,
\slashed{n}\bigg]\tilde{\Delta}_\sp(K)\Delta(K-P)\nn\\
&=&2g^2 C_F \sumintf_{\{ K\} } e^{-\frac{k_\perp^2}{q_fB}}k^0\tilde{\Delta}_\sp(K)\Delta(K-P).
\eea
From the above four expressions it can be noted that $b=-c$ and $d=-a$. These form factors will be calculated in Appendix~\ref{ff_sfa}.

\subsection{Effective propagator and dispersion relation}
The transverse momentum of the fermion becomes zero, i.e., $P_\perp=0$, in LLL. Thus the effective fermion propagator can be written using the Dyson-Schwinger equation as
\begin{align}
    S_{\text{eff}}(P_{\sp})=\frac{1}{\slashed{P_{\sp}}+\Sigma}.
\end{align}
Subsequently the inverse fermion propagator can be written as 
\begin{align}
    S^{-1}_{\text{eff}}(P)&={\slashed{P_{\sp}}+\Sigma}\\
    &=\mathcal{P}_R\slashed{L}\mathcal{P}_L+\mathcal{P}_L\slashed{R}\mathcal{P}_R,
\end{align}
where
\begin{align}
    &\slashed{L}=\slashed{P}+(a+c)\slashedl{u}+(b+d)\slashedl{n},\\
    &\slashed{R}=\slashed{P}+(a-c)\slashedl{u}+(b-d)\slashedl{n}.
\end{align}
Now the effective propagator can be written as
\begin{align}
    {S}_{\text{eff}}(P_{\sp})=\mathcal{P}_R\frac{\slashed{R}}{R^2}\mathcal{P}_L+\mathcal{P}_L\frac{\slashed{L}}{L^2}\mathcal{P}_R. \label{fermion_eff_S}
\end{align}
We have
\begin{align}
    &L^2=(p_0+(a+c))^2-\big(p_3-(b+d)\big)^2,\\
    &R^2=(p_0+(a-c))^2-\big(p_3-(b-d)\big)^2.
\end{align}
Now putting $a=-d$ and $b=-c$, one gets
\begin{align}
    &L^2=p_0^2-p_3^2 +2(a-b)(p_0-p_3) = (p_0-p_3) (p_0+p_3+2(a-b)),\\
    &R^2=p_0^2-p_3^2 +2(a+b)(p_0+p_3) = (p_0+p_3)(p_0-p_3+2(a+b)) .
\end{align}
Various discrete symmetries of the effective two-point functions are discussed in detail in Ref.~\cite{Das:2017vfh}.
The form factors are calculated in Appendix \ref{ff_sfa} and given as
\bea
a=-d=- \frac{g^2 C_F }{4\pi^2} \bigg[\sum_f q_f B\frac{p_0}{p_0^2-p_3^2} \ln 2 -\sum_{f}\(q_{f} B\)^2 
\frac{\zeta^\prime(-2)}{2T^2} \frac{p_0(p_0^2+p_3^2)}{(p_0^2-p_3^2)^2}\bigg],
\eea

\bea
b&=&-c
=\frac{g^2 C_F }{4 \pi^2}\bigg[ \sum_f q_f B\frac{p_3}{p_0^2-p_3^2} \ln 2  -\sum_f q_f B\frac{p_3}{2 T^2} \zeta^\prime(-2) \nn\\
&& - \sum_{f}\(q_{f} B\)^2   \frac{\zeta^\prime(-2)}{T^2} \frac{p_0^2 p_3}{(p_0^2-p_3^2)^2}\bigg].
\eea
The magnetic mass is found by taking the dynamic limit of $R^2$ and $L^2$ in  Eq.~(\ref{fermion_eff_S}), i.e., 
$R^2|_{p\rightarrow0,p_0=0}=L^2|_{p\rightarrow0,p_0=0},$ and is given by
\bea
M_{\text{sfa}}^2= \frac{g^2 C_F}{4\pi^2T^2}\left(\sum_f (q_f B)~T^2 \ln{4}-\sum_{f}\(q_{f} B\)^2  \zeta'(-2)\right).
\eea
One can notice that the magnetic mass is dependent on both magnetic field and temperature.

\begin{center}
\begin{figure}[tbh]
 \begin{center}
 \includegraphics[scale=0.7]{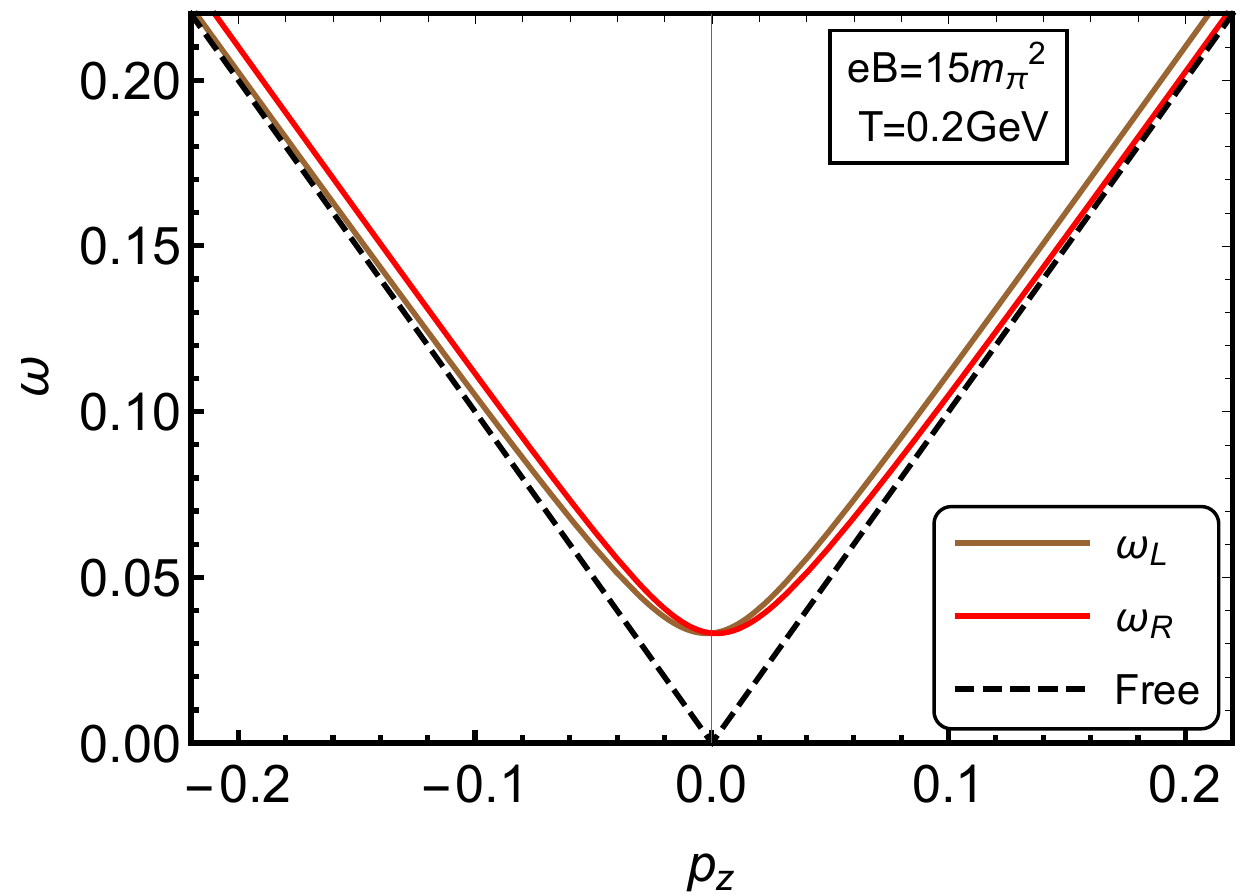} 
 \caption{Dispersion relation of the fermion in the presence of a strong magnetic field}
  \label{quark_disp}
 \end{center}
\end{figure}
\end{center}

Now we discuss the dispersion properties of fermions in a hot magnetized medium. 
The dispersion curves are obtained by solving, $L^2=0$ and $R^2=0$ given in Eq.~(\ref{fermion_eff_S}), numerically. There are four modes, two come from $L^2=0$ and two from $R^2=0$. In LLL only two modes are allowed~\cite{Das:2017vfh}: one $L$ mode with energy $\omega_L$ of a positively charged fermion having spin up and another one from $R$ mode with energy $\omega_R$ of a negatively charged fermion having spin-down. These two modes are plotted in Fig.~\ref{quark_disp}. In the LLL aproximation the transverse momentum of fermion becomes zero. Thus the dynamics of the system becomes two dimensional. At high $p_z$ both the modes of dispersion resembles the free dispersion mode. We also note that the reflection symmetry is broken in the presence of a magnetic field~\cite{Das:2017vfh}.

\subsection{One-loop quark free energy in the presence of a strongly magnetized medium}
\label{one_loop_quark_fe}
The quark free energy can be written as
\bea
F_q=- d_F \sumintf_{\{p_0\}}~\frac{d^3p}{(2\pi)^3}\ln{\left(\det[ S^{-1}_{\text{eff}}(p_0, p_3)]\right)} .
\label{fe}
\eea
Effective fermion self-energy can be written as
\bea
 S^{-1}_{\text{eff}}=\slashed{P_{\sp}}+\Sigma &=& (p_0+a)\slashedl{u}+(b-p_3)\slashedl{n}+c \gamma_5 \slashedl{u}+d \gamma_5 \slashedl{n}\nn\\
&=&(p_0+a)\gamma^0+(b-p_3)\gamma^3+c \gamma_5 \gamma^0+d \gamma_5 \gamma^3.
\eea
Now we evaluate the determinant as
\bea
\det[ S^{-1}_{\text{eff}}]&=&\bigg((b+c-p_3)^2-(a+d+p_0)^2\bigg)\bigg((-b+c+p_3)^2-(a-d+p_0)^2\bigg)\nn\\
&=&(p_0^2-p_3^2)\bigg((p_0+2a)^2-(p_3-2b)^2\bigg)\nn\\
&=& P_{\sp}^2\(P_\shortparallel^2+4 a p_0+4 b p_3+4 a^2- 4b^2\)\nn\\
&=&  P_{\sp}^4 \left(1+\frac{4a^2-4b^2+4ap_0+4bp_3}{P_{\sp}^2}\right),
\eea
where we have used $d=-a$ and $c=-b$. 

So Eq.~(\ref{fe}) becomes
\bea
F_q&=&- d_F \sumintf_{\{p_0\}}~\frac{d^3p}{(2\pi)^3} \ln\bigg[P_{\sp}^4\(1+\frac{4a^2-4b^2+4ap_0+4bp_3}{P_{\sp}^2}\)\bigg]\nn\\
&=& -2 d_F  \sumintf_{\{p_0\}}~\frac{d^3p}{(2\pi)^3}\ln{(-P_{\sp}^2)}- d_F \sumintf_{\{p_0\}}~\frac{d^3p}{(2\pi)^3}
\ln\bigg[1+\frac{4a^2-4b^2+4ap_0+4bp_3}{P_{\sp}^2}\bigg]\nn\\
&=& F^{\text{ideal}}_q+ F'_q,
\eea
where the free energy of free quarks~\cite{Strickland:2012vu}
\bea
F^{\text{ideal}}_q&=& -2 d_F  \sumintf_{\{p_0\}}~\frac{d^3p}{(2\pi)^3}\ln{(-P_{\sp}^2)}
= -2 d_F \sum_f \frac{q_f B}{(2 \pi)^2} \sumintf_{\{p_0\}} dp_3 \ln{(-P_{\sp}^2)}\nn\\
&=&- d_F \sum_f \frac{q_f B T^2}{12},
\eea
and
\bea
F'_q &=&- d_F \sumintf_{\{p_0\}}~\frac{d^3p}{(2\pi)^3}
\ln\bigg[1+\frac{4a^2-4b^2+4ap_0+4bp_3}{P_{\sp}^2}\bigg]\nn\\ 
&=&- d_F \sumintf_{\{p_0\}}~\frac{d^3p}{(2\pi)^3}\bigg[\frac{4\(a p_0 +b p_3\)}{P_{\sp}^2}+
\frac{4\(a^2 P^2-b^2P^2-2a^2 p_0^2-2b^2 p_3^2-4ab p_0p_3\)}{P_{\sp}^4}\nn\\
&&\hspace{4cm}+ \mathcal{O}(g^6)\bigg]\label{F_2_exp},
\eea
where we have kept terms up to $\mathcal{O}(g^4)$ to obtain the analytic exprssion of free energy. The expansion made above is valid for $g^2 (q_fB/T^2) < 1$, which can be
realized as  $(q_fB)/T^2 \gtrsim 1$ and $g \ll 1$.

As in the strong field approximation, the fermion is considered to be in LLL. So Eq.~(\ref{F_2_exp}) becomes,
\bea
F'_q&=& - d_F \sum_f\frac{q_f B}{(2\pi)^2}\sumintf_{\{p_0\}}~dp_3~\bigg[\frac{4\(a p_0 +b p_3\)}{P_{\sp}^2}+
\frac{4\(a^2 P^2-b^2P^2-2a^2 p_0^2-2b^2 p_3^2-4ab p_0p_3\)}{P_{\sp}^4}\nn\\
&&\hspace{4.5cm}+ \mathcal{O}(g^6)\bigg].
\eea
The sum-integrals are calculated in Appendix (\ref{quark_free_energy}), and the expression for the quark free energy up to $\mathcal{O}(g^4)$ is given by
\bea
F_q &=& - d_F \sum_f \frac{q_f B T^2}{12}-4d_F \sum_f \frac{\(q_f B\)^2}{(2\pi)^2}\frac{g^2 C_F}{4 \pi^2}\(\frac{\Lambda}{4\pi T}\)^{2\eps}\Bigg[\frac{1}{8\eps}  \(4\ln{2} - 
q_fB\frac{\zeta^{'} (-2)}{T^2} \)\nn\\
&&+\frac{1}{24576}\Bigg\{12288 \ln{2} (3 \gamma_E + 4\ln{2}-\ln{\pi})+\frac{256\zeta[3]}{\pi^4T^2}\Big(2 \pi^4 T^2 -3 g^2 C_F (q_fB) \ln{2}\nn\\
&&+ 
 3 \pi^2 (q_fB)(2 + 3 \gamma_E + 4\ln{2}- \ln{\pi})\Big)-\frac{8g^2C_F}{\pi^6T^4}(q_FB)^2 \zeta[3]^2(4 + 105 \ln{2})\nn\\
 &&+\frac{7245g^2C_F}{\pi^8T^6}(q_FB)^3
 \zeta[3]^3\Bigg\}\Bigg]\nn\\
&=&- d_F \sum_f\frac{q_f B T^2}{12} -4d_F \sum_f \frac{\(q_f B\)^2}{(2\pi)^2}\frac{g^2 C_F}{4 \pi^2}\Bigg[\frac{1}{8\eps}  \(4\ln{2} - 
q_fB\frac{\zeta^{'} (-2)}{T^2} \)+\frac{1}{24576}\nn\\
&&\Bigg\{12288 \ln 2\(3 \gamma_E +2 \ln \hat \Lambda + \ln 4 -\ln \pi\)+\frac{256\zeta[3]}{\pi^4T^2}\bigg(-3C_F g^2 q_fB \ln 2\nn\\
&&+6 \pi ^2 q_fB \ln \hat \Lambda +3 \pi ^2 q_fB (2+3 \gamma_E +\ln 4-\ln \pi )+2 \pi ^4 T^2\bigg)-\frac{8g^2C_F}{\pi^6T^4}(q_fB)^2\nn\\
&&\times \zeta[3]^2(4 + 105 \ln{2})+\frac{7245g^2C_F}{\pi^8T^6}(q_fB)^3
 \zeta[3]^3\Bigg\}\Bigg], \label{quark_hard}
 \eea
where $\hat \Lambda=\Lambda/2\pi T$. The quark free energy has ${\mathcal O}[\(q_f B\)^2/\epsilon]$ and ${\mathcal O}[\(q_f B\)^3/T^2\epsilon]$  divergences.


\section{Gluons  in a strong  magnetic field}
\label{GF}

\subsection{General structure of gauge boson free energy}
\label{ggb}

 The partition function for a gluon can generally be written in Euclidean space~\cite{Bandyopadhyay:2017cle} as
\be 
{\mathcal Z}_g = {\cal Z}  {\cal Z}^{\textrm{ghost}},~~
{\mathcal Z} = N_\xi\prod_{n,\bm{p}} \sqrt{\frac{(2\pi)^D}{\det D_{\mn, E}^{-1}}},~~
{\mathcal Z}^{\textrm{ghost}} = \prod_{n,\bm{p}} P_E^2,
\ee
where the product over $\bm{p}$ is for the spatial momentum whereas that over 
 $n$ is for the discrete bosonic Matsubara frequencies 
($\omega_n=2\pi n\beta;\, \, n=0,1,2,\cdots $) due to Euclidean time, $D$ is the spacetime 
dimension of the theory.  $D_{\mn,E}^{-1}$ is the inverse gauge boson propagator in Euclidean 
space with $P_E^2=\omega_n^2+p^2$ the square of four momentum.
$N_\xi=1/(2\pi\xi)^{D/2}$ is the normalization that originates from the introduction of a Gaussian integral 
at each location of position while averaging over the gauge condition function 
with a width $\xi$, the gauge fixing parameter. Gluon free energy can now be written~\cite{Bandyopadhyay:2017cle} as
\be
F_g = -d_A\frac{T}{V} \ln {\cal Z}_g = d_A\left[\frac{1}{2}\sumintb_{P_E}~\ln\Big[\textsf{det}
\left(D_{\mn, E}^{-1}(P_E)\right)\Big] -\sumintb_{P_E}~\ln P_E^2\right]. \label{fe_qed}
\ee
We note that   the presence  of the normalization factor $N_\xi$ eliminates  the gauge dependence explicitly.

For an ideal case $\textsf{det}\left(D_{\mn,E}^{-1}(P)\right)=(P_E^2)^4/\xi$, and hence the free energy for $d_A$ massless spin one gluons yields as
\bea
F_g^{\textrm{ideal}} = d_A\sumintb_{P_E}~\ln P_E^2 =d_A\sumintb_{P}~\ln\(- P^2\)= -d_A \frac{\pi^2T^4}{45},
\eea
where $P$ is four-momentum in Minkowski space and can be written as $P^2=p_0^2-p^2.$

\noindent In the presence of thermal background medium~\cite{Lebellac;1996,Kapusta:1989tk,Karmakar:2018aig} one can have
\bea
\textsf{det}\left(D_{\mn,E}^{-1}(P_E)\right) = \frac{P_E^2}{\xi}\left(P_E^2 + \Pi_T\right)^2\left(P_E^2 + \Pi_L\right),
\label{det_tqed}
\eea
which has four eigenvalues. Those are, respectively, $P_E^2/\xi$, $(P_E^2 + \Pi_L)$, and twofold degenerate 
$(P_E^2 + \Pi_T)$ where $\Pi_T$ and $\Pi_L$, respectively, are the transverse and longitudinal part of the gluon self-energy in a heat bath. 
Also we considered $D=4$, and  the spatial dimension, $d= 3$ 
throughout this manuscript\footnote{ We will also use $d=3-2\epsilon$ for 
dimensional regularization.}. From now on, we use Minkowski momentum $P$. Eventually the free energy 
becomes~\cite{Bandyopadhyay:2017cle}
\bea
F_g^{\textrm{th}} &=& \frac{1}{2}\left[\sumintb_{P}~\ln\(-P^2\) + 2\sumintb_{P}~\ln \left(-P^2 + \Pi_T\right) 
+ \sumintb_{P}~\ln \left(-P^2 + \Pi_L\right)\right] - \sumintb_{P}~\ln \(-P^2\) ,\nn\\
&=& \sumintb_{P}~\ln \left(-P^2+\Pi_T\right) + \frac{1}{2}\sumintb_{P}~\ln \left(1-\frac{\Pi_L}{P^2}\right) \\
&=& d_A\left[(d-1)F_g^T+F_g^L\right], \label{thfq}
\eea
Also, $F_g^L$ and $F_g^T$ are, respectively, the longitudinal and transverse part of the gluon free energy. Using
general structure of two-0point functions of the gauge boson, both of them are evaluated in Refs.~\cite{Lebellac;1996,Kapusta:1989tk,Karmakar:2018aig}. 

 Now, the general structure of the inverse propagator of a gauge boson in ther presence of a hot magnetized medium 
 is computed in Ref.~\cite{Karmakar:2018aig}  as
\bea
\left(\mathcal{D}_{\mn}\right)^{-1} = \frac{P^2}{\xi}\eta_{\mn} + 
\left(P_m^2 - b\right)B_{\mn} + \left(P_m^2 - c\right)R_{\mn} + 
\left(P_m^2 - d\right)Q_{\mn}, 
\label{inverse_prop}
\eea
where 
\bea
P_m^2 = P^2\frac{\xi -1}{\xi}
\eea
and $b$, $c$, $d$ are the form factors corresponding to the three projection tensors $B^{\mn}$,$R^{\mn}$ 
and $Q^{\mn}$, respectively for gauge boson self-energy~\cite{Karmakar:2018aig}.
The determinant of the inverse of the gauge boson propagator can be evaluated from Eq.~\eqref{inverse_prop} as
\bea
\textsf{det}\left(D_{\mn,E}^{-1}(P)\right) = -\frac{P^2}{\xi}\left(-P^2+b\right)\left(-P^2
+c\right)\left(-P^2+d\right), \label{det_mqed} 
\eea
which has four eigenvalues: $-P^2/\xi, \, \left(-P^2+b\right), \, \left(-P^2+c\right), \, {\mbox{and}} \, 
\left(-P^2+d\right)$. We note here that one has two distinct transverse modes coming from $(-P^2 + c)=0$ and $(-P^2 + d)=0$ respectively, in a thermomagnetic medium  instead of  a two fold degenerate transverse mode $(-P^2 +\Pi_T)=0$ in thermal medium in Eq.~\eqref{det_tqed}. 

Using Eq.~\eqref{det_mqed} in Eq.~\eqref{fe_qed}, the one-loop gluon free energy for hot magnetized medium is given~\cite{Karmakar:2018aig}  by
\bea
F_g = d_A
\left[\mathcal{F}_g^1+\mathcal{F}_g^2+\mathcal{F}_g^3\right], 
\label{free_qed}
\eea
where
\begin{subequations}
\begin{align}
\mathcal{F}_g^1 &= \frac{1}{2}\sumintb_{P}~\ln\left(1-\frac{b}{P^2}\right), \label{f1_qed}\\
\mathcal{F}_g^2 &= \frac{1}{2}\sumintb_{P} ~\ln\left(-P^2 + c\right),\label{f2_qed}\\
\mathcal{F}_g^3 &= \frac{1}{2}\sumintb_{P} ~\ln\left(-P^2 + d\right).\label{f3_qed}
\end{align}
\end{subequations}
The various structure functions are obtained in Ref.~\cite{Karmakar:2018aig} in both strong and weak field approximation. In the following sections we obtain the gluon free energy in strong field approximation.
\subsection{Gluon free energy  in a strongly magnetized hot and dense medium}
\label{GFS}
\begin{center}
\begin{figure}[tbh]
 \begin{center}
 \includegraphics[scale=0.4]{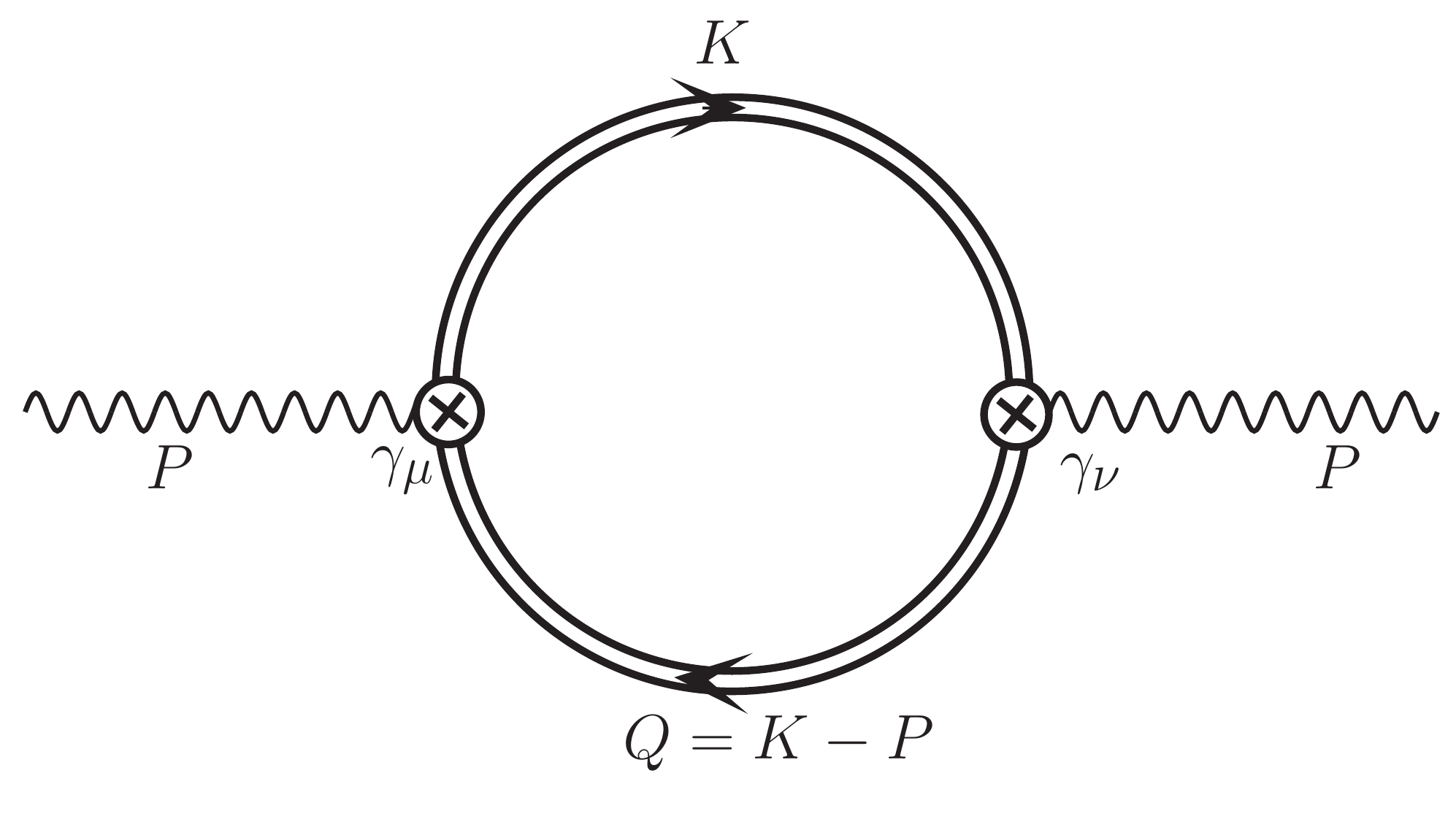} 
 \caption{Self-energy diagram for a gluon in the strong magnetic field approximation. 
The double line indicates the modified quark propagator in the presence of a strong magnetic 
field.}
  \label{gluon_se}
 \end{center}
\end{figure}
\end{center} 
Within the strong field approximation ($m_f^2< T^2 < q_fB$) the expressions for the different contributions 
of the gluon free energy  in LLL can now be obtained from Eqs.~(\ref{free_qed}).
Combining  Eqs.~\eqref{f1_qed}, \eqref{f2_qed}, and~\eqref{f3_qed} with  Eq.~\eqref{free_qed}, the total one-loop free energy  expanded up to ${\mathcal O}[g^4]$ 
is given 
by
\bea
F_g 
&\approx&d_A\left[\sumintb_{P} ~\ln\left(-P^2\right) -\frac{b+c+d}{2P^2}-\frac{b^2+c^2+d^2}{4P^4}\right], \label{Fsg_expan}
\eea 
where the expansion is made to obtain analytical expression for free energy which  is valid for $g^2 (q_fB/T^2) < 1$, and it can be realized as 
$(q_fB)/T^2 \gtrsim 1$ and $g \ll 1$.

Now the various structure functions involved therein are obtained in Ref.~\cite{Karmakar:2018aig}. For 
simplicity we consider $m_f=0$ with a hierarchy of scales as $m_f^2\sim m^2_{\textrm{th}} \sim g^2T^2 < T^2 < q_fB$ and write 
down those structure functions~\cite{Karmakar:2018aig} as 
\begin{subequations}
\begin{align}
b &=\frac{C_Ag^2T^2}{3\bar u^2}\left[1-\mathcal{T}_P(p_0,p)\right]-\sum_f \frac{g^2 q_fB}{4\pi^2\bar u^2}e^{{-p_\perp^2}/{2q_fB}}~\frac{p_3^2}{p_0^2-p_3^2}, \label{b_sf} \\
c&=\frac{C_Ag^2T^2}{3}\frac{1}{2}\left[\frac{p_0^2}{p^2}-\frac{P^2}{p^2}\mathcal{T}_P(p_0,p)\right] , \label{c_sf} \\
d&= \frac{C_Ag^2T^2}{3}\frac{1}{2}\left[\frac{p_0^2}{p^2}-\frac{P^2}{p^2}\mathcal{T}_P(p_0,p)\right]+\sum_f \frac{g^2 q_fB}{4\pi^2\bar u^2} e^{{-p_\perp^2}/{2q_fB}}~ \frac{p_3^2}{p_0^2-p_3^2}, \label{d_sf}
\end{align}
\end{subequations}
where ${\bar u}^2 = - p^2/P^2$, $\mathcal{T}_P(p_0,p)=\frac{p_0}{2p}\ln\frac{p_0+p}{p_0+p}$.

Now we write down various terms in Eq.~(\ref{Fsg_expan}) as
\bea
\sumintb_P \,\,\frac{b+c+d}{2P^2}&=& \frac{C_A g^2T^2}{6}\sumintb_P \frac{1}{P^2}\nn\\
\sumintb_P \,\,\frac{b^2+c^2+d^2}{4P^4}&=& \(\frac{C_A g^2T^2}{3}\)^2\frac{1}{4}\sumintb_P \Bigg\{\frac{3\mathcal{T}_p^2}{2p^4}
+\frac{1}{2P^4}+\frac{1}{p^2P^2}-\frac{3\mathcal{T}_p}{p^4}-\frac{\mathcal{T}_p}{p^2P^2}\Bigg\} \nn\\
&+&\frac{1}{2}\sum_{f_1,f_2} \(\frac{g^2B}{4\pi^2}\)^2q_{f_1}q_{f_2} \sumintb_P\,\, e^{-\frac{p_{\perp}^2}{2q_{f_1}B}}
e^{-\frac{p_{\perp}^2}{2q_{f_2}B}}\frac{p_3^4}{p^4\(p_0^2-p_3^2\)^2}\nn\\
&+&\sum_f \frac{g^2q_fB}{4\pi^2}\frac{C_A g^2T^2}{3} \sumintb_P\,
e^{-\frac{p_{\perp}^2}{2q_fB}} \frac{p_3^2}{4\(p_0^2-p_3^2\)}\(\frac{3\mathcal{T}_p}{p^4}-\frac{3}{p^4}-\frac{1}{p^2P^2}\).
\eea
\noindent In the strong field limit Eq.~\eqref{Fsg_expan} becomes 
\bea
F_g&=&d_A\Bigg[\sumintb_{P} ~\ln\left(-P^2\right)-\frac{C_A g^2T^2}{6}\sumintb_P \frac{1}{P^2}-\(\frac{C_A g^2T^2}{3}\)^2\frac{1}{4}\sumintb_P \Bigg\{\frac{3\mathcal{T}_p^2}{2p^4}+\frac{1}{2P^4}\nn\\
&&+\frac{1}{p^2P^2}-\frac{3\mathcal{T}_p}{p^4}-\frac{\mathcal{T}_p}{p^2P^2}\Bigg\}-\frac{1}{2}\sum_{f_1,f_2} \(\frac{g^2B}{4\pi^2}\)^2q_{f_1}q_{f_2}  \sumintb_P\,\, e^{-\frac{p_{\perp}^2}{2q_{f_1}B}}e^{-\frac{p_{\perp}^2}{2q_{f_2}B}}\nn\\
&&\times\frac{p_3^4}{p^4\(p_0^2-p_3^2\)^2}-\sum_f \frac{g^2q_fB}{4\pi^2}\frac{C_A g^2T^2}{3} \sumintb_P\, e^{-\frac{p_{\perp}^2}{2q_fB}} \frac{p_3^2}{4\(p_0^2-p_3^2\)}\Big(\frac{3\mathcal{T}_p}{p^4}-\frac{3}{p^4}\nn\\
&&-\frac{1}{p^2P^2}\Big)\Bigg].\label{fe_sfa}
\eea
The first term in Eq.~\eqref{fe_sfa} gives us the free case.  Using the sum-integrals listed in Eqs.~(\ref{gluon_1}) -(\ref{gluon_9}), the hard contribution of the one-loop gluon free energy in a strongly magnetized hot medium is calculated in Appendix \ref{HAS} and can be written as
 \bea
F_g^{\text{hard}}&=& \frac{d_A}{(4\pi)^2}\Bigg[\frac{1}{\eps}\Bigg\{-\frac{1}{8}\(\frac{C_A g^2T^2}{3}\)^2+\frac{g^4T^4}{96}\sum_{f_1,f_2} \frac{q_{f_1}B}{q_{f_2}B} +\frac{N_f^2g^4T^4}{96}+ \frac{C_A N_fg^4T^4}{36} \nn\\ 	&-&\sum_{f_1,f_2} \frac{g^4(q_{f_1}B)(q_{f_2}B)}{64\pi^4}+N_f\sum_{f} \frac{g^4T^2q_{f}B}{32\pi^2}-\sum_f \frac{1}{4\pi^2}\frac{C_A g^4T^2q_fB}{6}\(1+\ln 2\)\Bigg\} \nn\\
 	&-& \frac{16 \pi^4 T^4}{45}+\frac{2C_A g^2\pi^2T^4}{9}+\frac{1}{12}\(\frac{C_A g^2T^2}{3}\)^2 \left(8 - 3 \gamma_E - \pi^2 + 4\ln 2 - 3 \ln\frac{\hat \Lambda}{2}\right)\nn\\
 	&+&
 	\frac{N_f\pi ^2 T^2}{2} \(\frac{g^2}{4\pi^2}\)^2  \sum_{f} q_fB\bigg( \frac{2\zeta'(-1)}{\zeta(-1)}-1+2\ln\hat \Lambda\bigg) +\left(N_f^2+\sum_{f_1,f_2}\frac{q_{f_1}B}{q_{f_2}B}
 	\right)\nn\\
 	&\times&	\frac{g^4T^4}{32} \left(\frac{2}{3}\ln\frac{\hat \Lambda}{2} -\frac{60  \zeta '[4]}{\pi^4}-\frac{1}{18} (25-12 \gamma_E -12 \ln 4 \pi ) \right) -\frac{1}{2}\(\frac{g^2}{4\pi^2}\)^2\nn\\
 	&\times&	\sum_{f_1,f_2} q_{f_1}B q_{f_2}B \bigg( \ln\frac{\hat \Lambda}{2}+\gamma_E +\ln 2\bigg)
 	- \frac{C_AN_f g^4T^4}{36} \bigg(1 - 2\frac{\zeta'(-1)}{\zeta(-1)} - 2 \ln\frac{\hat \Lambda}{2}\bigg)\nn\\
 	&-&\sum_f \frac{C_Ag^4T^2q_fB}{144\pi^2} \bigg(\pi ^2-4+12\ln \frac{\hat \Lambda }{2}-  2\ln 2\(6\gamma_E+4+3\ln2-6\ln \frac{\hat \Lambda }{2}\) +12 \gamma_E  \bigg)\Bigg].\nn\\ \label{gluon_hard}
 	\eea

As it can be seen, $F_g^{\text{hard}}$ has an ${\cal O}(1/\epsilon)$ divergence from the HTL approximation as well as from the thermomagnetic contribution. 

We get the soft contribution of gluon free energy by considering soft gluon momentum $(P \sim gT)$  with  $p_0=0$, 
\bea
F_g^{\text{soft}}&\approx&d_A\bigg[-\frac{(m_D^s)^3T}{12\pi}+\mathcal O[\eps]\bigg], \label{gluon_soft}
\eea
where the Debye mass in a strong field~\cite{Bandyopadhyay:2017cle} is given by
\bea
({m_D^s})^2=\frac{g^2N_cT^2}{3}+\sum_f \frac{g^2 q_fB}{4\pi^2}. \label{mass_debye}
\eea
The total gluonic contribution becomes 
\be
F_g=F_g^{\text{hard}}+F_g^{\text{soft}}. \label{gluon_fe}
\ee
\section{Anisotropic pressure of deconfined QCD matter in a strong magnetic field}
\label{thermo}
\subsection{Renormalized free energy in a strong field approximation}
Combining Eq.~\eqref{total_fe} and Eq.~\eqref{gluon_fe} one-loop free energy of deconfined QCD matter 
in the presence of a strong magnetic field can be written as
\bea
F&=&F_q+F^{\text{hard}}_g+F_g^{\text{soft}}+F_0 + \Delta {\mathcal E}^0_T +\Delta {\mathcal E}_T^B, \label{fe_unrenor}
\eea
which has ${\mathcal O}[1/\epsilon]$ divergences in various orders $(q_fB)$. The
$\mathcal O[(q_fB)^2]$ divergences present in the free energy are regulated by redefining the tree level free energy $B^2/2$ as
\bea
F_0= \frac{B^2}{2}&\rightarrow& \frac{B^2}{2}
\underbrace{+4d_F \sum_f \frac{(q_fB)^2}{(2\pi)^2}\frac{g^2 C_F}{4 \pi^2}\frac{\ln{2}}{2\eps} 
+\frac{d_A}{(4\pi)^2}\sum_{f_1,f_2} \frac{g^4q_{f_1}Bq_{f_2}B}{64\pi^4 \eps}}_{\Delta {\mathcal E}^{B^2}} \nn \\
&\rightarrow & \frac{B^2}{2} + {\Delta {\mathcal E}^{B^2}}. \label{div_B2}
\eea
Now the other divergences of $\mathcal O[ (q_fB)^0T^4]$, $\mathcal O[T^2 (q_fB)]$, and $\mathcal O[(q_fB)^3/T^2]$  are renormalized by adding suitable counterterms as follows: the $\mathcal O[ (q_fB)^0T^4]$ divergences are regulated through counterterms as
\bea
\Delta {\mathcal E}^0_T&=& \Delta {\mathcal E}^{\mbox{\tiny{HTL}}}_T+\Delta {\mathcal E}_T\nn\\
&=& \underbrace{d_A\frac{m_D^4}{128\pi^2 \eps}}_{\Delta {\mathcal E}^{\mbox{\tiny{HTL}}}_T} 
\underbrace{-\frac{d_A}{(4\pi)^2}\left[ \frac{g^4T^4}
{96 \eps}\sum_{f_1,f_2} \frac{q_{f_1}B}{q_{f_2}B}+\frac{N_f^2g^4T^4}{96 \eps}+\frac{C_A N_f g^4T^4}{36 \eps}\right]}_{\Delta {\mathcal E}_T}, \label{div_B0}
\eea
where $m_D$ is the Debye screening mass in the HTL approximation. Now the $\mathcal O[T^2 (q_fB)]$ and $\mathcal O[(q_fB)^3/T^2]$ divergences are regulated through counterterms
\bea
\Delta {\mathcal E}_T^B&=&-4d_F \sum_f \frac{\(q_f B\)^3}{(2\pi)^2}\frac{g^2 C_F}{4 \pi^2} 
\frac{\zeta^{'} (-2)}{8T^2\eps}-\frac{d_A}{(4\pi)^2\eps}\Bigg[ \frac{N_f g^4T^2}{32\pi^2}\sum_{f}q_fB\nn\\
&&-\sum_f \frac{1}{4\pi^2}\frac{C_A g^4T^2q_fB}{6}\(1+\ln 2\)\Bigg].  \label{div_B13}
\eea

Now using Eqs.~\eqref{quark_hard}, \eqref{gluon_hard}, \eqref{gluon_soft}, \eqref{div_B2}, \eqref{div_B0} and \eqref{div_B13} in  Eq.~\eqref{fe_unrenor}, one 
obtains renormalized one-loop quark-gluon free energy in the presence of a strong magnetic field as
\bea
F=F^r_q+F^r_g + \frac{B^2}{2} , \label{fe_renor}
\eea
where
renormalized quark free energy $F_q^r$ is given by
\bea
F_q^r&=&- d_F \sum_f\frac{q_f B T^2}{12} -4d_F \sum_f \frac{\(q_f B\)^2}{(2\pi)^2}\frac{g^2 C_F}{4 \pi^2}\Bigg[\frac{1}{24576}\Bigg\{12288 \ln 2\big(3 \gamma_E\nn\\
&&+2 \ln \hat \Lambda 
+ \ln 16\pi\big)+\frac{256\zeta[3]}{\pi^4T^2}\bigg(-3C_F g^2 q_fB \ln 2+6 \pi ^2 q_fB \ln \hat \Lambda +3 \pi ^2 q_fB \nn\\
&\times&(2+3 \gamma_E +\ln 16\pi )+2 \pi ^4 T^2\bigg)-\frac{8g^2C_F}{\pi^6T^4}(q_fB)^2 \zeta[3]^2(4 + 105 \ln{2})\nn\\
&&+\frac{7245g^2C_F}{\pi^8T^6}(q_fB)^3
 \zeta[3]^3\Bigg\}\Bigg], \label{fq}
 \eea
 and the renormalized total gluon free energy containing both hard and soft contributions is given as
 \bea
F_g^r&=& \frac{d_A}{(4\pi)^2}\Bigg[
 	- \frac{16 \pi^4 T^4}{45}+\frac{2C_A g^2\pi^2T^4}{9}+\frac{1}{12}\(\frac{C_A g^2T^2}{3}\)^2 \bigg(8 - 3 \gamma_E - \pi^2 + 4\ln 2 \nn\\
 	&&- 3 \ln\frac{\hat \Lambda}{2}\bigg)
 	+
 	\frac{N_f\pi ^2 T^2}{2} \(\frac{g^2}{4\pi^2}\)^2  \sum_{f} q_fB\bigg( \frac{2\zeta'(-1)}{\zeta(-1)}-1+2\ln\hat \Lambda\bigg)\nn\\
 	&&+\bigg(N_f^2+\sum_{f_1,f_2}\frac{q_{f_1}B}{q_{f_2}B}
 	\bigg)
    \frac{g^4T^4}{32} \left(\frac{2}{3}\ln\frac{\hat \Lambda}{2} -\frac{60  \zeta '[4]}{\pi^4}-\frac{1}{18} (25-12 \gamma_E -12 \ln 4 \pi ) \right) \nn\\
    &&-\frac{1}{2}\(\frac{g^2}{4\pi^2}\)^2
 		\sum_{f_1,f_2} q_{f_1}B q_{f_2}B \bigg( \ln\frac{\hat \Lambda}{2}+\gamma_E +\ln 2\bigg)
 	- \frac{C_AN_f g^4T^4}{36} \bigg(1 - 2\frac{\zeta'(-1)}{\zeta(-1)} \nn\\
 	&&- 2 \ln\frac{\hat \Lambda}{2}\bigg)-\sum_f \frac{C_Ag^4T^2q_fB}{144\pi^2} \bigg(\pi ^2-4+12\ln \frac{\hat \Lambda }{2}-  2\ln 2\bigg(6\gamma_E+4+3\ln2\nn\\
 	&&-6\ln \frac{\hat \Lambda }{2}\bigg) +12 \gamma_E  \bigg)\Bigg]-\frac{d_A( m_D^s)^3T}{12\pi}.\label{fg}
 	\eea

\subsection{Longitudinal and transverse pressures}
In the thermal background one can calculate  QCD pressure from the free energy of the system and the pressure is isotropic. Now in the presence of thermo-magnetic background, one has another extensive parameter as external magnetic field $B$. In this case free energy can be written as
\bea
\mathcal F(T,V,B)=E^{\text{total}}-TS-eB\cdot \mathcal{M},\nn
\eea
where ${\cal M}$ is the magnetization.
Free energy density in a  finite spatial volume $V$ is given by 
\bea
F=\mathcal{F}/V=\eps^{\text{total}}-Ts-eB\cdot M,
\eea
where $\eps^{\text{total}}$ is the total energy density and, the entropy density is given by
\bea
s=-\frac{\partial F}{\partial T}\label{entropy}, 
\eea
and the magnetization per unit volume is given by
\bea
M=-\frac{\partial F}{\partial (eB)} \label{magnetization}
\eea
and the total energy density $\eps^{\text{total}}=\eps+\eps^{\text{field}}$. $\eps$ is the energy density of the medium and $\eps^{\text{field}}=eB\cdot M$. In the presence of a strong magnetic field the space becomes anisotropic, and one gets different pressures~\cite{PerezMartinez:2007kw} for directions parallel and perpendicular to the magnetic field. Longitudinal and transverse pressures are given as 
\bea
P_{z}=-F ,\,\,\, P_{\perp}=-F-eB\cdot M=P_z-eB\cdot M.\label{trans_pressure}
\eea

\subsection{Pressure of ideal quark and gluon  gas in a strong magnetic field}
\label{ideal_fe}
The free energy of an ideal quark-gluon gas in the absence of a magnetic field is given as
\bea
F_T^{\text{ideal}}=-d_F\frac{7\pi^2 T^4}{180}-d_A\frac{\pi^2 T^4}{45} \label{F_id0},
\eea
and the corresponding pressure reads as
\bea
P_T^{\text{i}}\equiv P_T^{\text{ideal}}&=&d_F\frac{7\pi^2 T^4}{180}+d_A\frac{\pi^2 T^4}{45} \label{F_id0}\nn\\
&\equiv& (P_T^q)^{\text{i}} +(P_T^g)^{\text{i}}.\label{P_id_0}
\eea
The free energy of an ideal quark-gluon gas in the presence of a magnetic field is given by
\bea
F^{\text{ideal}}&=&F_q^{\text{ideal}}+F_g^{\text{ideal}}\nn\\
&=&- d_F \sum_f (q_f B) \frac{ T^2}{12}-d_A\frac{ \pi^2 T^4}{45}.
\eea
As seen the  quarks are affected by the magnetic field whereas the electric charge less gluons are not affected by the magnetic field. The quark contribution in the presence of a strong magnetic field  makes the ideal quark-gluon gas  pressure  anisotropic. The ideal longitudinal pressure is given by
\bea
P_z^{\text{i}}\equiv P_z^{\text{ideal}}&=&-F^{\text{ideal}}\nn\\
&=&  d_F \sum_f (q_f B) \frac{ T^2}{12}+d_A\frac{ \pi^2 T^4}{45}\nn\\
&\equiv&(P_z^q)^{\text{i}} +(P_z^g)^{\text{i}}. \label{P_id_w_B}
\eea
Magnetization of the ideal quark-gluon gas is calculated using Eq.~\eqref{magnetization} as
\bea
M^{\text{ideal}}&=&-\frac{\partial F^{\text{ideal}}}{\partial (eB)}\nn\\
&=&  d_F \sum_f\frac{  q_f T^2}{12}.
\eea
As found, the magnetization of a  ideal quark-gluon gas in LLL in the presence of a strong magnetic field is independent of the magnetic field. In LLL positive charge particles with spin-up align along the magnetic field  direction, whereas negative charge particles with spin-down align opposite to the magnetic field direction. Because of this the system remains in a minimum free energy configuration with respect to $eB$. Now even if one increases the magnetic field the spin alignment does not change; thus for given a $T$ the ideal quark-gluon gas acquires a constant magnetization. However, if one increases $T$, the spin alignment in LLL again does not change but the increased  thermal motion  along the field direction can  result in an increase of magnetization.
 
Now the ideal transverse pressure of the quark-gluon gas can be written using Eq.~\eqref{trans_pressure} as
\bea
P_\perp^{\text{i}}\equiv P_{\perp}^{\text{ideal}}&=& d_A\frac{ \pi^2 T^4}{45}\label{P_ideal_perp}
\eea
We note that the transverse pressure of ideal magnetized quark-gluon gas  is independent of the magnetic field
and is the same as the ideal gluon pressure.  As discussed above the  gluons are not affected by the magnetic field and contribute to this isotropic pressure.  On the other hand quarks have momenta only along the $z$ direction in LLL and contribute only to the  longitudinal pressure.
\section{Results}
\label{res}
We use one-loop running coupling constant that evolves on both the momentum transfer and the magnetic field~\cite{Ayala:2018wux} as
\bea
\alpha_s(\Lambda^2, |eB|)&=&\frac{\alpha_s(\Lambda^2)}{1+b_1\, \alpha_s(\Lambda^2)
\ln\left(\frac{{\Lambda^2}}{{\Lambda^2\ +\ |eB|}}\right)},
\eea
in the strong magnetic field domain $|eB| > \Lambda^2$. The one-loop running coupling in the absence of a magnetic field at the renormalization scale is given as
\bea
\alpha_s(\Lambda^2)&=&\frac{1}
{b_1\, \ln\left({\Lambda^2}/{\Lambda_{\overline{{\rm MS}}}^2}\right)},
\eea
where $b_1= {(11N_c-2N_f)}/{12\pi}$ and $\Lambda_{\overline{{\rm MS}}}=176~{\rm MeV}$ \cite{Beringer:1900zz} at 
$\alpha_s(1.5\, {\mbox{GeV}})=0.326 $ for $N_f=3$. The renormalization scale is chosen as $\Lambda=2\pi T$. The renormalization scale can be varied by a factor of $2$  with respect to its central value. Furthermore, we are interested in the thermomagnetic correction here and hence we will drop the tree level vacuum
contribution $B^2/2$ from our discussion. We also note that some orders of coupling $g$ are not complete in one-loop HTL calculation. To have a complete picture of pressure 
up to a specific order of $g$, one needs to perform higher loop order calculation. However, as a first effort we confine ourselves  in one loop calculation here.

\begin{center}
\begin{figure}[tbh]
 \begin{center}
 \includegraphics[scale=0.57]{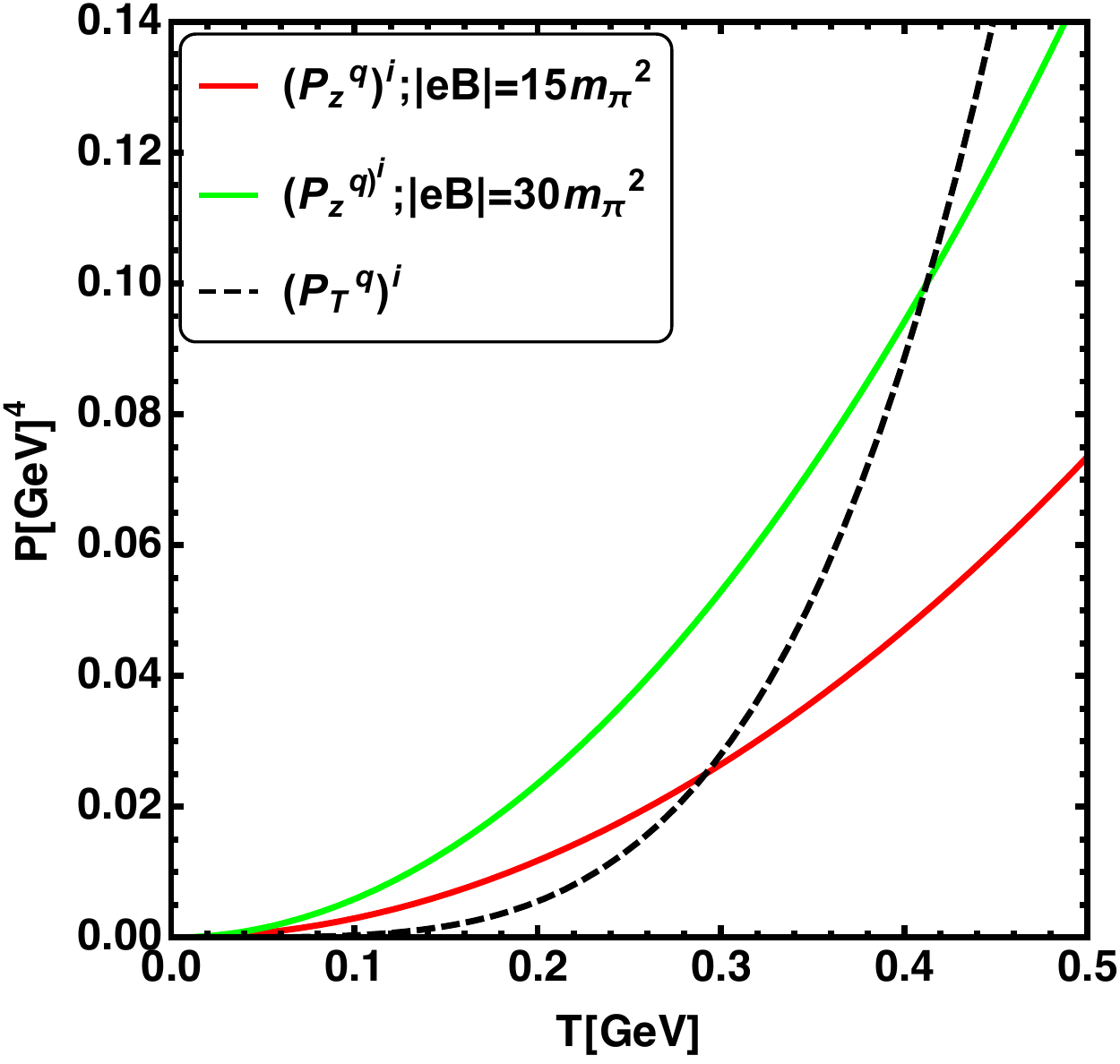}
 \caption{ Variation of ideal quark pressure with and without a magnetic field as a function of temperature.}
  \label{quark_T}
 \end{center}
\end{figure}
\end{center}
For ideal quark-gluon gas the gluons remain unaffected but quarks are strongly affected in the presence of a magnetic field. So in Fig.~\ref{quark_T} we display a variation of the ideal quark pressure with [($P_z^q)^{\text{i}}$  in Eq.~\eqref{P_id_w_B}]  and without [($P_T^q)^{\text{i}}$ in Eq.~\eqref{P_id_0}] magnetic field as a function of temperature. The ideal quark pressure,$(P_z^q)^{\text{i}}$ , in the presence of a magnetic field is proportional to $(eB)T^2$ whereas that in the absence of a magnetic field, $(P_T^q)^{\text{i}}$, is proportional to $T^4$. For a given magnetic field,  $T^2$ dominates at low T whereas  $T^4$ dominates at high $T$, and  thus a crossing takes place at an intermediate temperature as seen in Fig.~\ref{quark_T}. Also the ideal longitudinal pressure increases linearly  with the increase of the magnetic field as can be seen from  Eq.~\eqref{P_id_w_B}.

\begin{center}
\begin{figure}[tbh]
 \begin{center}
 \includegraphics[scale=0.5]{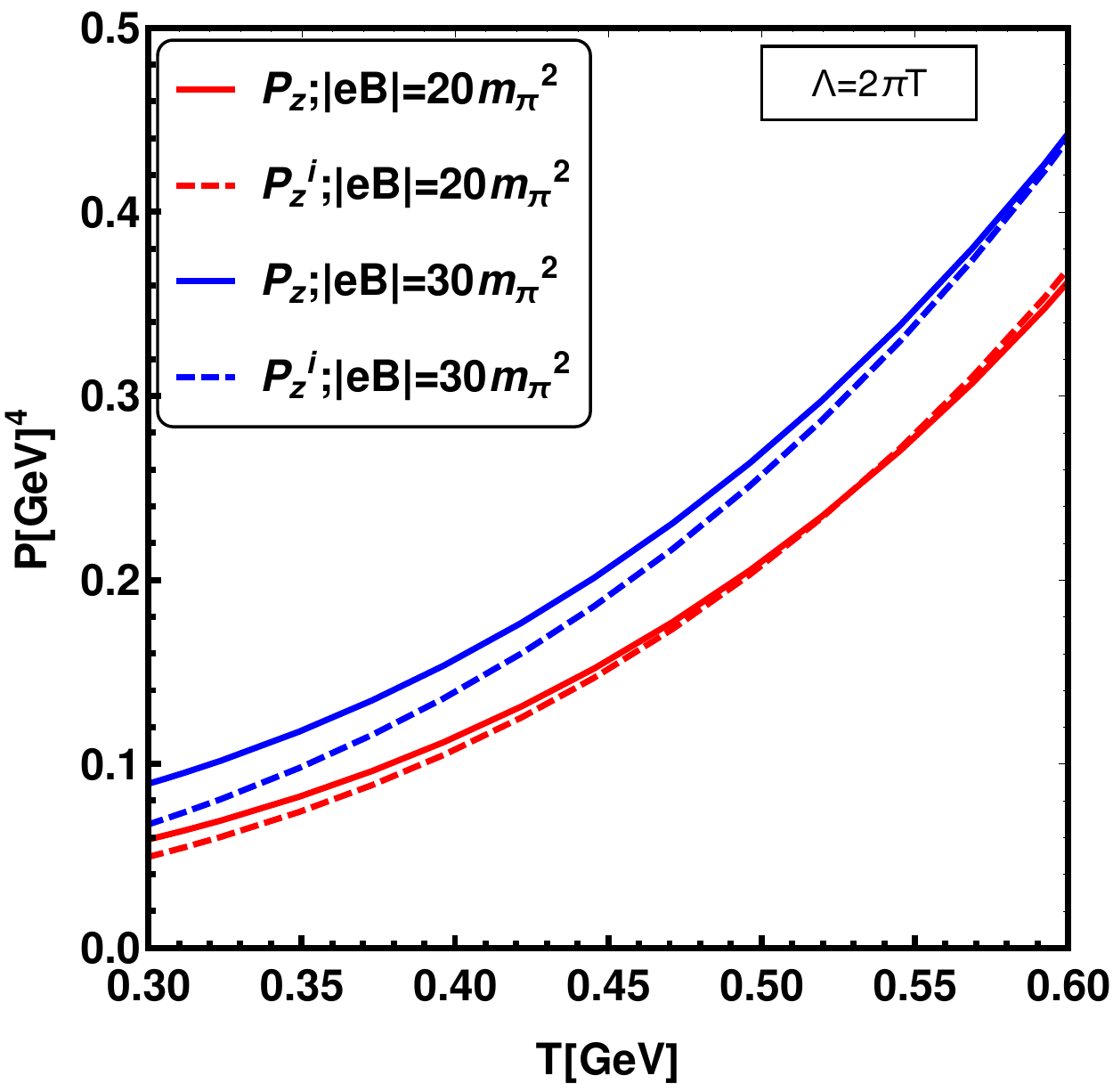}
  \includegraphics[scale=0.5]{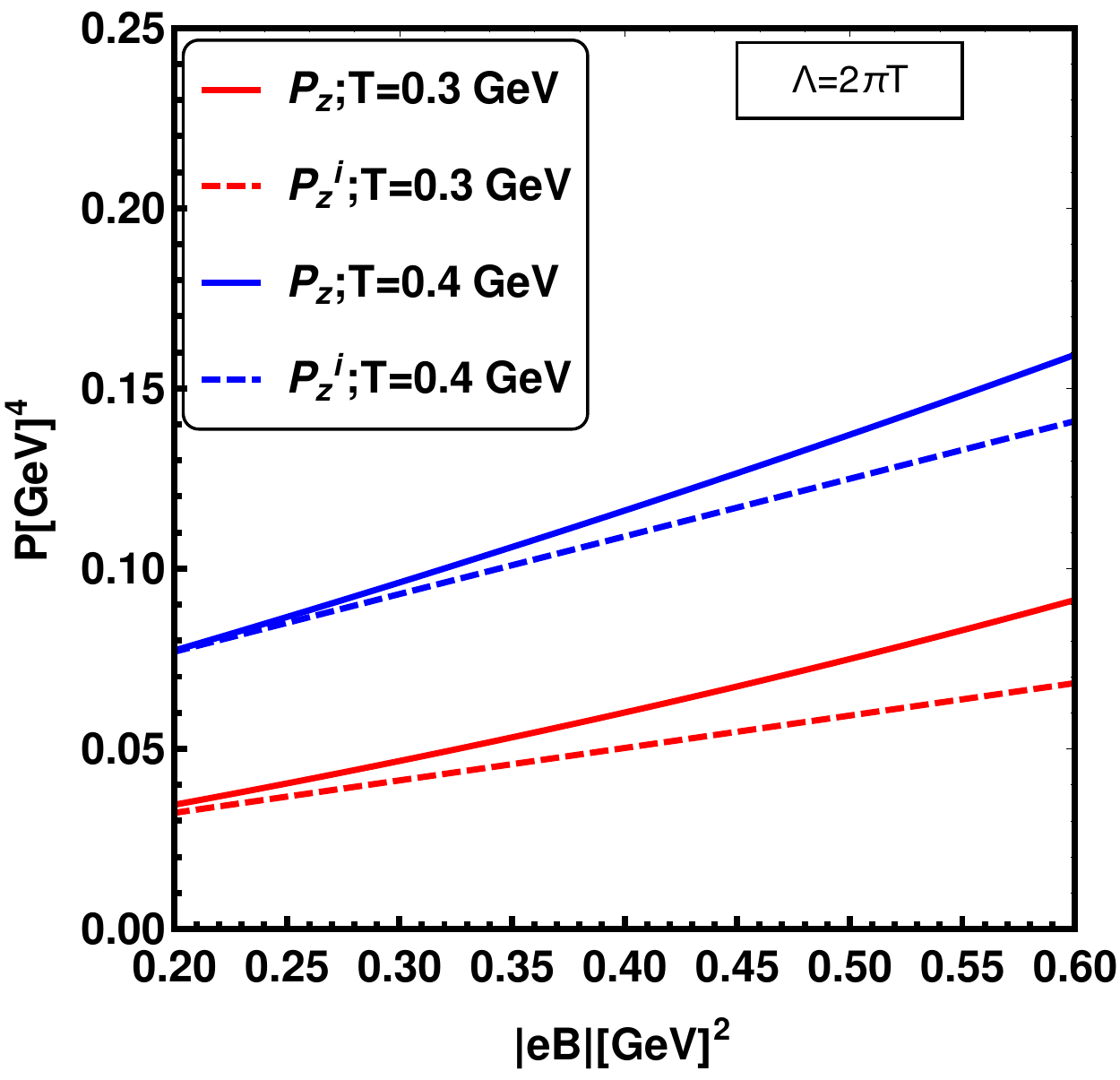}
 \caption{Variation of one-loop longitudinal pressure as a function of temperature for different values of the magnetic field (left panel) and as a function of the magnetic field at different temperatures (right panel) for $N_f=3$ and the central value of the  renormalization scale, $\Lambda=2\pi T$. Dashed curves represent ideal longitudinal pressure.}
  \label{1loop_long_pressure_T_eB}
 \end{center}
\end{figure}
\end{center}

\begin{center}
\begin{figure}[tbh]
 \begin{center}
 \includegraphics[scale=0.54]{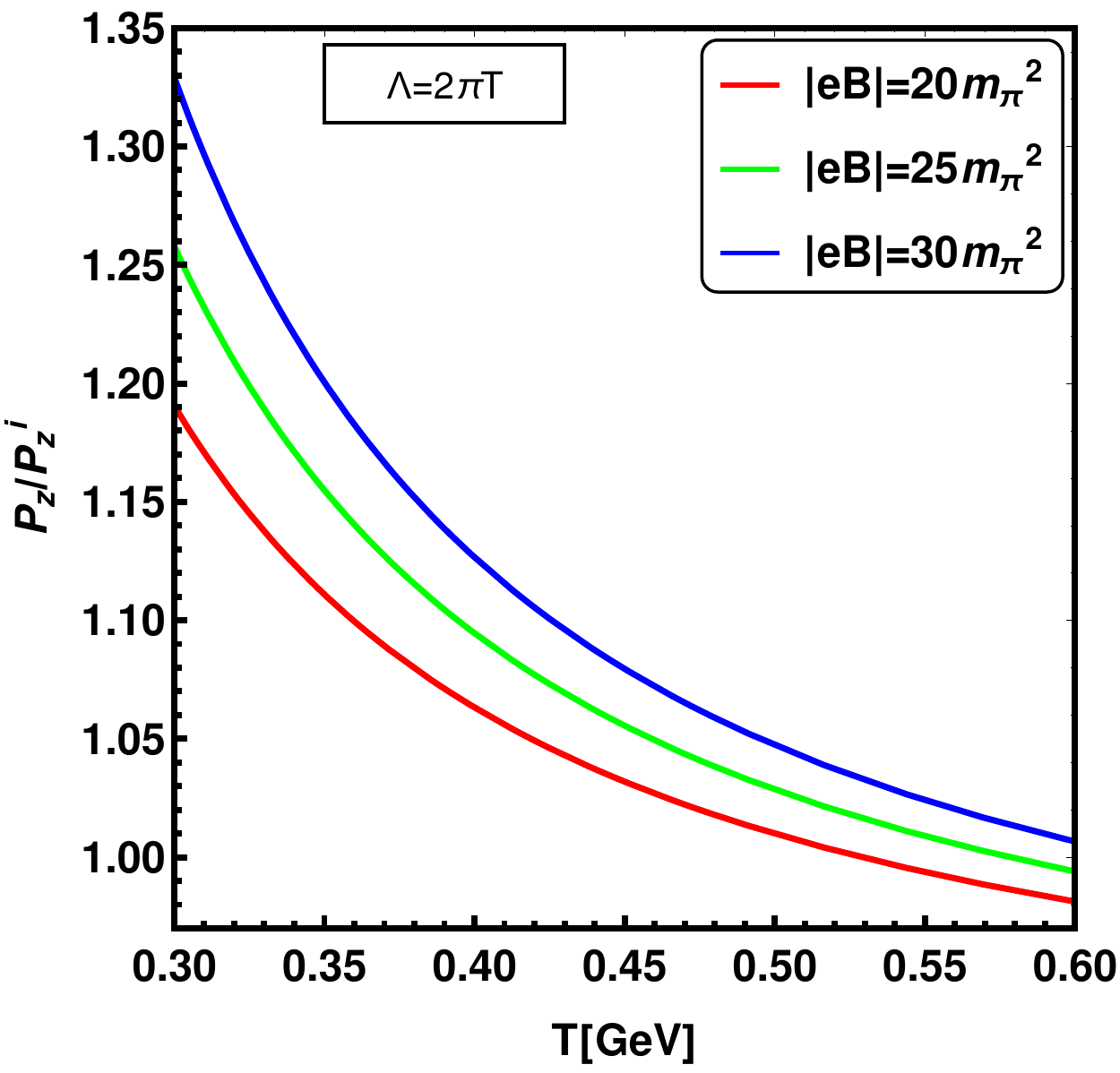}
   \includegraphics[scale=0.5]{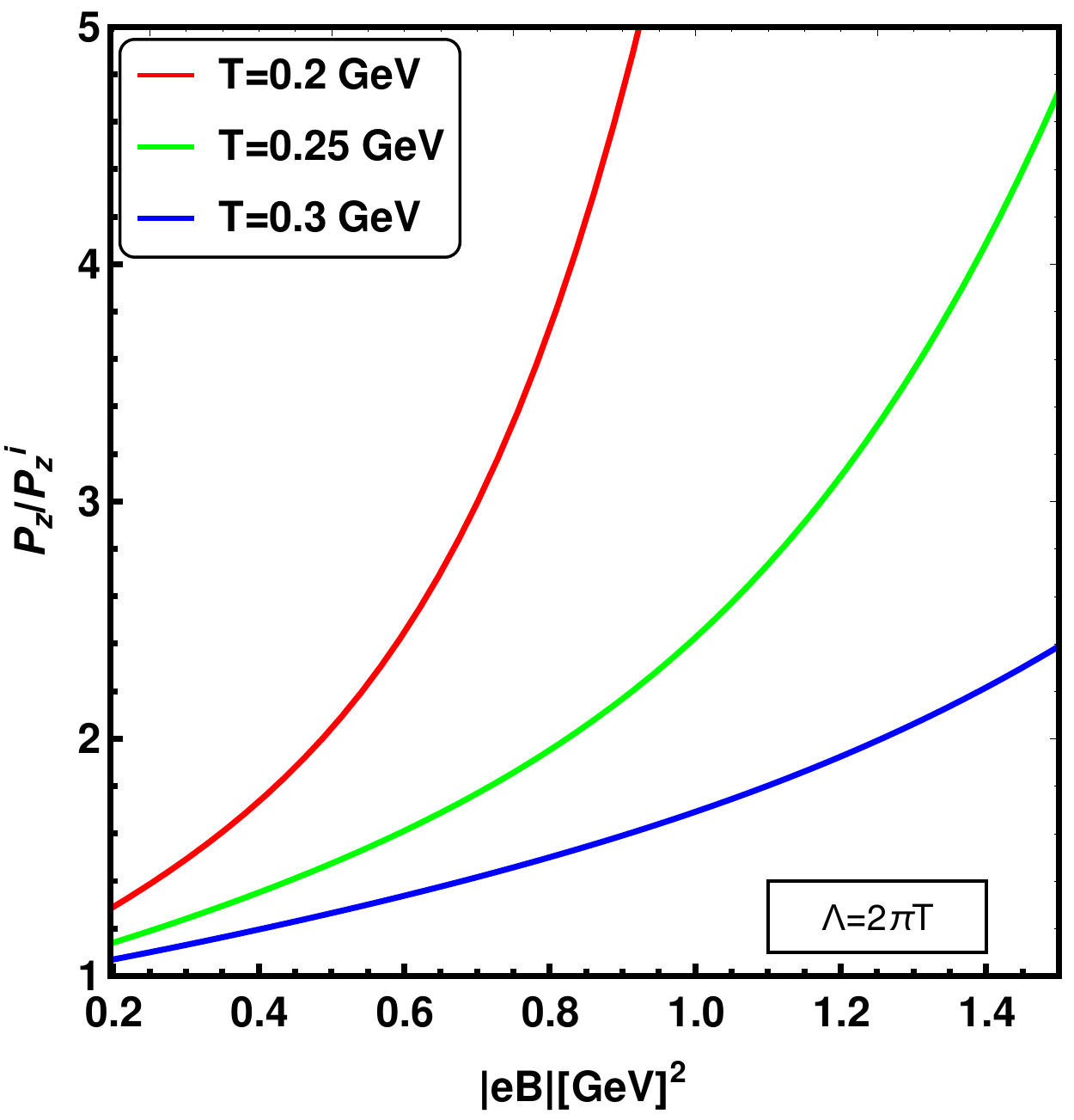}
 \caption{Variation of the one-loop longitudinal pressure scaled with the ideal longitudinal pressure as a function of temperature (left panel) for different values of the magnetic field and as a function of the magnetic field (right panel) for different temperatures with $N_f=3$. }
  \label{1loop_long_pressure_T_eB_scale}
 \end{center}
\end{figure}
\end{center}

The left panel of Fig.~\ref{1loop_long_pressure_T_eB} displays a comparison of one-loop longitudinal pressure (solid curve) and ideal pressure (dashed curve) with temperatures for different values of field strength, whereas the right panel displays the same but with the strength of the magnetic field for different temperatures. In both cases one-loop pressure increases with the increase in temperature and field strength, respectively. However, the one-loop interacting pressure is higher than that of the ideal one in both panels. This enhancement can be understood as follows: in one-loop order both the effective quark two-point function and the effective 
gluon two-point function containing a quark loop are strongly affected in the presence of the magnetic field, which contribute to the additional pressure compared to the ideal case. For a given magnetic field this enhancement is stronger in the temperature domain $(300-500)$ MeV as can be seen from the scaled pressure with with ideal one ($P_z/P_z^{\text{i}}$) in the left panel of Fig.~\ref{1loop_long_pressure_T_eB_scale}.  However, this enhancement gradually decreases with increase of temperature and approaches to ideal value at high temperature. For a given temperature the ratio ($P_z/P_z^{\text{i}}$)  increases with the increase of magnetic field strength as found in the right panel
 of Fig.~\ref{1loop_long_pressure_T_eB_scale}. This is because $P_z^{\text{i}}$ has linear dependence on $eB$ whereas $P_z$ has a higher power dependence on $eB$.

\begin{center}
\begin{figure}[tbh]
 \begin{center}
 \includegraphics[scale=0.57]{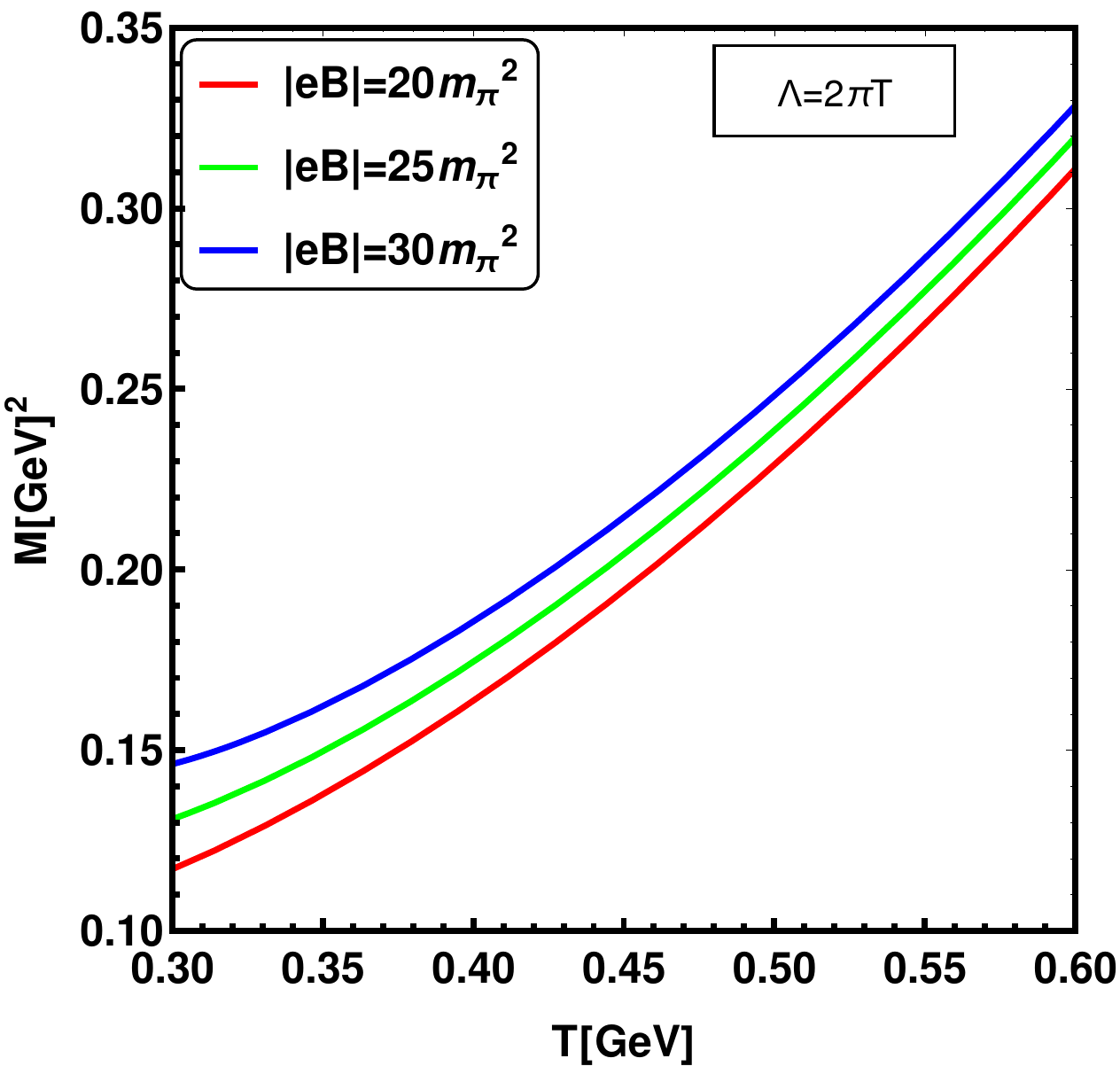}
  \includegraphics[scale=0.57]{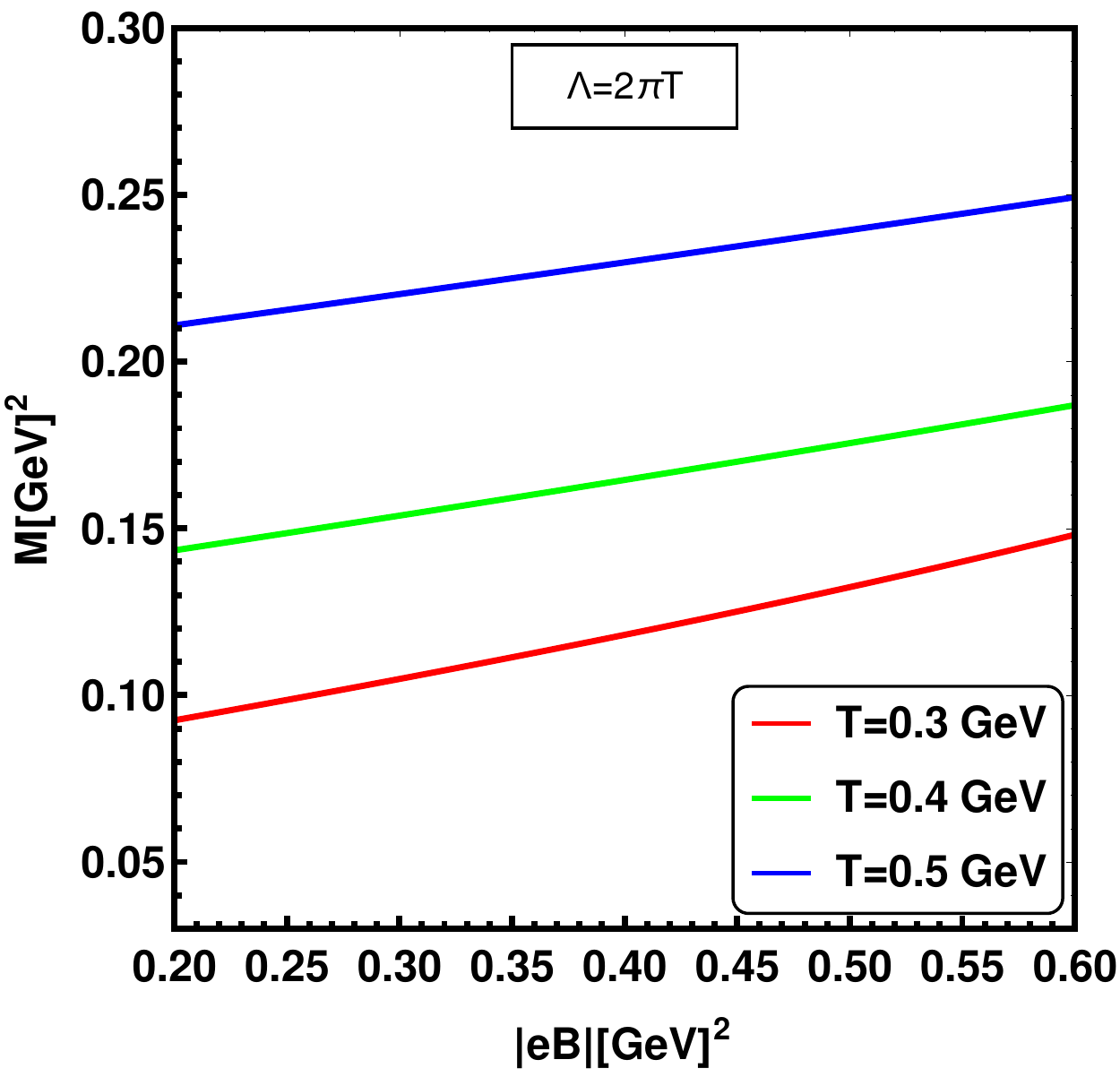}
 \caption{Variation of magnetization with temperatures for different magnetic fields  (left panel) and with magnetic fields for various temperatures (right panel).}
  \label{1loop_magneti}
 \end{center}
\end{figure}
\end{center}
The magnetization of an ideal quark-gluon gas in the presence of a magnetic field has already been discussed in Sec.~\ref{ideal_fe}. Now the magnetization of an interacting quark-gluon system is calculated using Eq.~\eqref{magnetization} and it is proportional to $[a T^2+b (eB)+c (eB)^2/T^2+d (eB)^3/T^4+f (eB)^4/T^6]$
which is plotted in Fig.~\ref{1loop_magneti}. So for a given value of $eB$, at low $T$ limit $1/T^n$ with $n=2,4,6$ terms dominate but are restricted by the scale $gT$, whereas at high $T$, $T^2$ terms dominate which is seen from the left panel. In contrast to the ideal quark-gluon gas the magnetization of an interacting quark-gluon system increases with the magnetic field that is displayed in the right panel.\footnote{Even if the fermions are in LLL, the magnetization increases with magnetic field due to interactions.} This trend is in agreement with LQCD results~\cite{Bali:2013owa}. In the strong magnetic field approximation $g^2T^2<T^2 < eB$, the magnetization acquires  positive values with range $0<M<1$. So the deconfined QCD matter in the presence of a strong magnetic field within one-loop HTL approximation shows a paramagnetic nature (i.e., magnetization is parallel to the field direction)~\cite{Bali:2013owa}. Since the magnetization in a strong field limit increases with the magnetic field, it causes an increase in pressure of the system along the field direction, i.e., longitudinal direction. This in turn also strongly affects the transverse pressure as we would see below.
 \begin{center}
\begin{figure}[tbh]
 \begin{center}
 \includegraphics[scale=0.57]{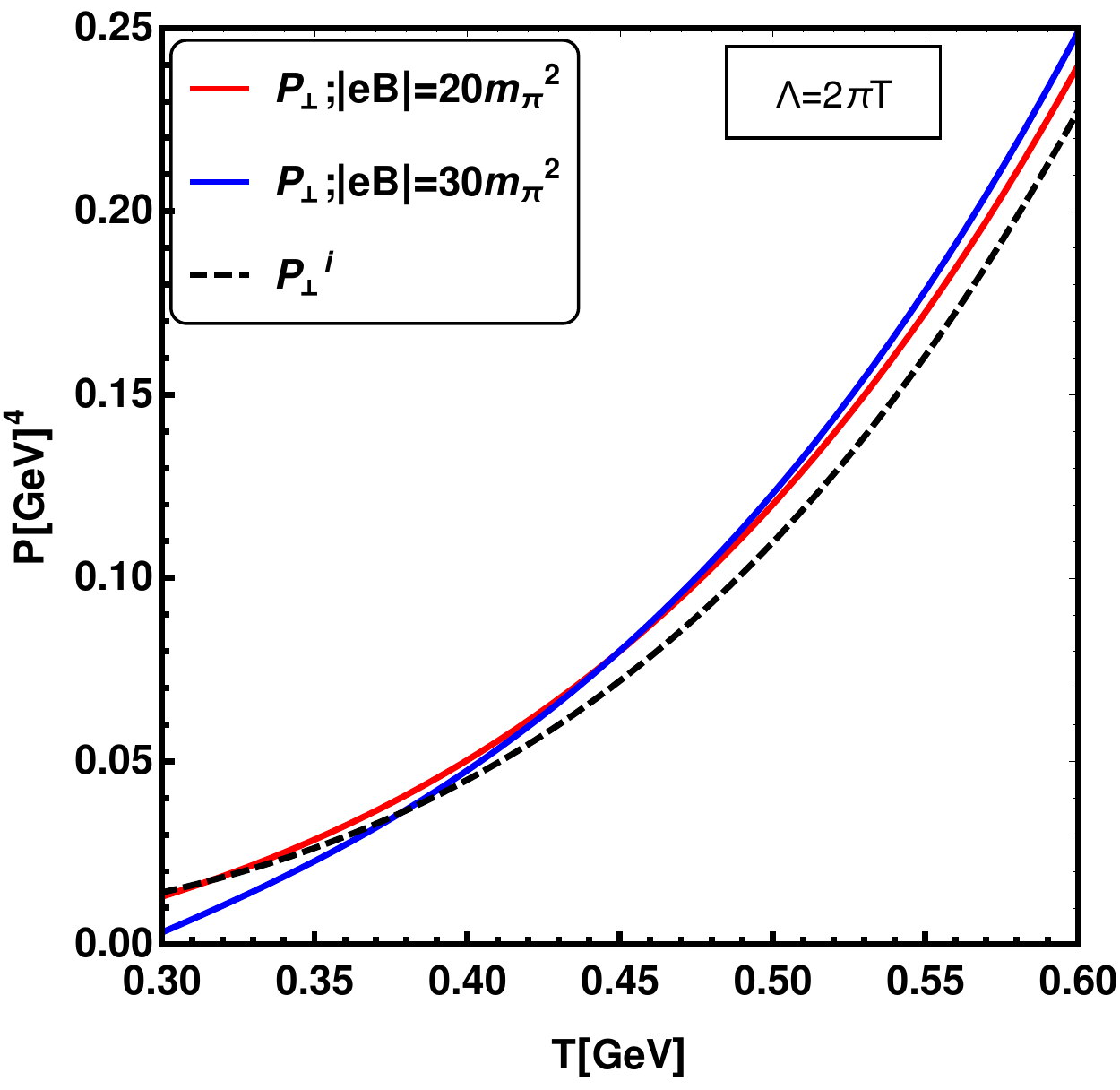}
    \includegraphics[scale=0.57]{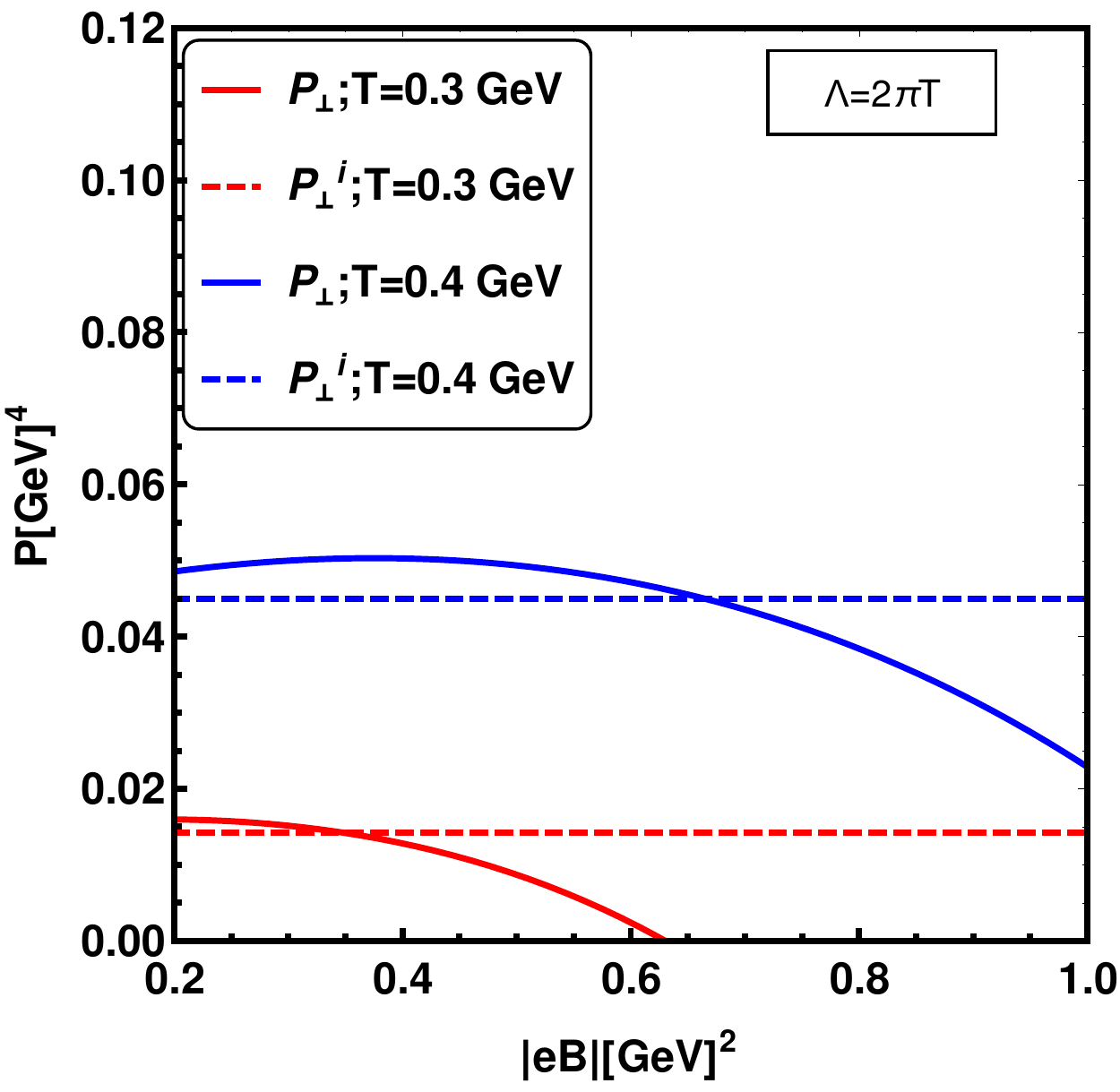}
 \caption{Variation of the one-loop transverse pressure as a function of temperature for various magnetic fields (left panel) and as a function of magnetic fields for different temperatures (right panel). Dashed curves represent the ideal transverse pressure.}
  \label{trans_ideal_unscaled_T_eB}
 \end{center}
\end{figure}
\end{center}
One-loop transverse pressure is calculated using Eq.~\eqref{trans_pressure}. It is evident from Eq.~\eqref{trans_pressure} and from the left panel of Fig.~\ref{trans_ideal_unscaled_T_eB} that the one-loop transverse pressure increases with temperature and shows a nature similar to longitudinal pressure (left panel of Fig.~\ref{1loop_long_pressure_T_eB}) but lower in magnitude. Dashed lines represent the transverse ideal pressure which is independent of a magnetic field as given in Eq.~\eqref{P_ideal_perp}. For a given high value of a magnetic field the pressure starts with a lower value than that of the ideal gas particularly at low $T$, and then a crossing takes place. This could also be understood from the right panel. In the right panel the transverse pressure is displayed as a function of the magnetic field for two different temperatures. Also the dashed lines here represent the ideal transverse pressure that is independent of the magnetic field.  The transverse pressure for the interacting case is given in Eq.~\eqref{trans_pressure} as $P_{\perp}=P_z-eB \cdot M$. Now for a given temperature its variation is very slow (or almost remain unaltered) with a lower value of the magnetic field because there is a competition between $P_z$ and $eBM$. Since the magnetization $M$ increases steadily with magnetic the field (right panel of Fig.~\ref{1loop_magneti}) the transverse pressure, $P_\perp$, tends to decrease, falls below the ideal gas value  and  may even go to negative values for low $T$ at a large value of the magnetic field. This is an indication that the system may shrink in the transverse direction~\cite{Bali:2013owa}.

\section{Conclusion}
\label{conclu}
We consider the deconfined QCD matter in the presence of the strong background magnetic field within the HTL approximation. Quarks are directly affected by the external magnetic field. In strong field approximation we assume the quarks are in the lowest Landau level. Gluons are affected through the quark loop in the gluon two-point function. Hard and soft contributions of 
quark-gluon free energy are calculated within the one-loop HTL approximation. Various divergent terms are eliminated by choosing appropriate counterterms in the $\overline{MS}$ renormalization scheme. In the presence of a strong magnetic field the hot QCD matter 
acquires paramagnetic nature. The presence of magnetization makes the system anisotropic and one gets different pressures in direction parallel and perpendicular to the magnetic field.
Both longitudinal (along magnetic field direction) and transverse (perpendicular to the magnetic field direction) pressures are evaluated completely analytically by calculating the magnetization of the system. Various thermodynamic properties can be studied using the obtained free 
energy here. Moreover, the anisotropic pressure obtained here may be useful for a magnetohydrodynamics  description and analysis of elliptic flow of hot and dense deconfined QCD matter created in heavy-ion collisions. Finally we note that since we have considered the one-loop HTL perturbation theory up to $\mathcal O[g^4]$, $\mathcal O[g^2]$ and $\mathcal O[g^4]$ are incomplete. The present result could be improved by going to higher loop orders. 

\section{Acknowledgments}
B.K., R.G., and M.G.M. were funded by Department of Atomic Energy (DAE), India via the 
project TPAES. A.B. acknowledges the support from the research grants from Conselho Nacional de
Desenvolvimento Cient\'{\i}fico e Tecnol\'ogico (CNPq), under grant Coordena\c c\~ao de
Aperfei\c coamento de Pessoal de N\'ivel Superior (CAPES), Govt of Brazil. N.H. was funded by DAE, India. B.K. acknowledges helpful discussions with Anwesha Chattopadhyay.
\appendix

\section{Calculation of form factors}
\label{ff_sfa}
Before computing the quark form factors, we first decompose the transverse and the longitudinal parts from the expression of the self-energy in Eq.~(\ref{self_sfa_tot}). Considering the transverse parts of the fermionic momenta to be relatively weaker in the strong field limit, we make an assumption that $(K-P)_\perp < (K-P)_\sp$. 
Then we can write down the gluonic propagator as 
\bea
\Delta(K-P) &=&  \frac{1}{(K-P)_\sp^2-(k-p)_\perp^2} = \frac{1}{(K-P)_\sp^2}\left[1-\frac{(k-p)_\perp^2}{(K-P)_\sp^2}\right]^{-1} \nn\\
&\approx& \frac{1}{(K-P)_\sp^2}\left(1+\frac{(k-p)_\perp^2}{(K-P)_\sp^2}\right)\nn\\
&=& \Delta_\sp(K-P)+ (k-p)_\perp^2\Delta_\sp^2(K-P)\label{quark_approx}.
\eea

\subsection{Calculation of quark form factor $a$}
We calculate the form factor $a$ by using Eqs.~\eqref{quark_a} and~\eqref{quark_approx} as
\bea
a=\frac{1}{4}\Tr[\Sigma \slashed u]&=&- 2g^2 C_F \sumintf_{\{K\}} e^{-\frac{k_\perp^2}{q_fB}}\left[\frac{k_0}{K_\sp^2(K-P)^2_\sp} + (k-p)_\perp^2\frac{k_0}{K_\sp^2 (K-P)^4_\sp}\right]\nn\\
&=& -2g^2 C_F  \int \frac{d^3k}{(2\pi)^3} e^{-\frac{k_\perp^2}{q_fB}}\left[T_2 + (k-p)_\perp^2 T_4\right]\nn\\
&=& -2g^2 C_F  \int_{-\infty}^{\infty} \frac{dk_3}{2\pi} \left[ \frac{q_fB}{4\pi} T_2 + \frac{q_f B}{4\pi}(p_{\perp}^2+q_fB)\hspace{.2cm} T_4\right]\nn\\
&=&- \frac{g^2 C_F (q_f B)}{4\pi^2}  \int_{-\infty}^{\infty} dk_3 \left[  T_2 + (p_{\perp}^2+q_fB)\hspace{.2cm} T_4\right],\label{a}
\eea
where
\bea
T_2&=& \sum \frac{k_0}{K_\sp^2(K-P)^2_\sp},\nn\\
T_4&=&  \sum \frac{k_0}{K_\sp^2 (K-P)^4_\sp}
= -\frac{1}{2k_3}\frac{\partial T_2}{\partial p_3}.
\eea
Here we also note that in LLL, $p_{\perp}=0$. Now
\bea
T_2&=& -\frac{1}{4q_3}\Bigg[\left(1+n_B(q_3)-n_F(k_3)\right)\left(\frac{1}{p_0+k_3+q_3}+\frac{1}{p_0-k_3-q_3}\right)\nn\\
&&+\big(n_B(q_3)+n_F(k_3)\big)\left(\frac{1}{p_0-k_3+q_3}+\frac{1}{p_0+k_3-q_3}\right)\Bigg]\nn\\
&\approx& -\frac{1}{4k_3}\Bigg[ \left(n_B(k_3)-p_3\frac{\partial n_B(k_3)}{\partial k_3}+n_F(k_3)\right) \left(\frac{1}{p_0-p_3}+\frac{1}{p_0+p_3}\right)\Bigg]\nn\\
&=& -\frac{1}{4k_3}\Bigg[n_F(k_3)+\left(n_B(k_3)-p_3\frac{\partial n_B(k_3)}{\partial k_3}\right)
\Bigg]\frac{2p_0}{p_0^2-p_3^2}.
\eea
\bea
\int_{-\infty}^{\infty} dk_3 \,T_2 &=&  \frac{2p_0}{p_0^2-p_3^2} \bigg[-\frac{1}{4}\int \frac{dk_3}{k_3}\left(n_F(k_3)+n_B(k_3)-p_3\frac{\partial n_B}{\partial k_3}\right)\bigg]\nn\\
&\approx& \frac{2p_0}{p_0^2-p_3^2}\bigg[-\frac{1}{2}\int_{0}^{\infty} dk_3\,\frac{n_F}{k_3}-\frac{1}{2}\int_{0}^{\infty} dk_3\,\frac{n_B}{k_3}\bigg]\nn\\
&=& \frac{p_0}{p_0^2-p_3^2}\bigg[\left(\frac{1}{4\eps}+\frac{\gamma_E}{2}+\frac{1}{2}\ln{\frac{2}{\pi}}\right)+\left(-\frac{1}{4\eps}-\frac{\gamma_E}{2}+\frac{1}{2}\ln{2\pi}\right)\bigg]+\mathcal{O}[\eps]\nn\\
&=&\frac{p_0}{p_0^2-p_3^2} \ln 2.
\eea
Similarly, we get
\bea
\int_{-\infty}^{\infty} dk_3\, T_4 &=&- \frac{1}{2} \frac{\partial}{\partial p_3} \int_{-\infty}^{\infty}  \frac{dk_3}{k_3}T_4\nn\\
&=&- \frac{\zeta^\prime(-2)}{2T^2}\frac{p_0(p_0^2+p_3^2)}{(p_0^2-p_3^2)^2}.
\eea
So
\bea
a=-d=- \frac{g^2 C_F (q_f B)}{4\pi^2} \bigg[ \frac{p_0}{p_0^2-p_3^2} \ln 2 -q_f B \frac{\zeta^\prime(-2)}{2T^2} \frac{p_0(p_0^2+p_3^2)}{(p_0^2-p_3^2)^2}\bigg].
\eea

\subsection{Calculation of quark form factor $b$}
Similarly, one can calculate $b$ from Eqs.~\eqref{quark_b} and~\eqref{quark_approx} as
\bea
b=-\frac{1}{4}\Tr[\Sigma \slashed n]&=& 2g^2 C_F \sumintf_{\{K\}} e^{-\frac{k_\perp^2}{q_fB}}\left[\frac{k_3}{K_\sp^2(K-P)^2_\sp} + (k-p)_\perp^2\frac{k_3}{K_\sp^2 (K-P)^4_\sp}\right]\nn\\
&=& 2g^2 C_F  \int \frac{d^3k}{(2\pi)^3} e^{-\frac{k_\perp^2}{q_fB}}\,k_3\,\left[T_1 + (k-p)_\perp^2 T_3\right]\nn\\
&=& 2g^2 C_F  \int_{-\infty}^{\infty} \frac{dk_3}{2\pi} \,k_3\,\left[ \frac{q_fB}{4\pi} T_1 + \frac{q_f B}{4\pi}(q_fB)\hspace{.2cm} T_3\right]\nn\\
&=& \frac{g^2 C_F (q_f B)}{4 \pi^2}  \int_{-\infty}^{\infty} dk_3 \hspace{.2cm}k_3 \left[  T_1 +q_fB\hspace{.2cm} T_3\right],
\label{b}
\eea
where
\bea
T_1&=& \sum \frac{1}{K_\sp^2(K-P)^2_\sp},\nn\\
T_3&=& \sum \frac{1}{K_\sp^2 (K-P)^4_\sp}
=- \frac{1}{2k_3}\frac{\partial T_1}{\partial p_3}.
\eea
Now
\bea
T_1&=& \frac{1}{4k_3 q_3}\Bigg[\left(1+n_B(q_3)-n_F(k_3)\right)\left(\frac{1}{p_0+k_3+q_3}-\frac{1}{p_0-k_3-q_3}\right)\nn\\
&&+\left(n_B(q_3)+n_F(k_3)\right)\left(\frac{1}{p_0+k_3-q_3}-\frac{1}{p_0-k_3+q_3}\right)\Bigg]\nn\\
&\approx& \frac{1}{4k_3^2}\Bigg[ \left(n_B(k_3)-p_3\frac{\partial n_B(k_3)}{\partial k_3}-n_F(k_3)\right) \frac{1}{k_3}+ \bigg(n_B(k_3)-p_3\frac{\partial n_B(k_3)}{\partial k_3}\nn\\
&&+n_F(k_3)\bigg)\left(\frac{1}{p_0+p_3}-\frac{1}{p_0-p_3}\right)\Bigg]\nn\\
&=&  \frac{1}{4k_3^2}\Bigg[ \left(n_B(k_3)-p_3\frac{\partial n_B(k_3)}{\partial k_3}-n_F(k_3)\right) \frac{1}{k_3}\nn\\
&&+ \left(n_B(k_3)-p_3\frac{\partial n_B(k_3)}{\partial k_3}+n_F(k_3)\right) \left(\frac{-2p_3}{p_0^2-p_3^2}\right)\Bigg].
\eea

\bea
 \mbox{Let\,\,\,}\int_{-\infty}^{\infty} dk_3 \hspace{.2cm} k_3\, T_1 &=& I_1 \, \, {\mbox{with}}
\eea
\bea
I_1&=& \int_{-\infty}^{\infty} dk_3 \frac{1}{4k_3}\Bigg[ \left(n_B(k_3)-p_3\frac{\partial n_B(k_3)}{\partial k_3}-n_F(k_3)\right) \frac{1}{k_3}\nn\\
&&~~~~~~~~~~~~~~~~~~~~~~~ +\left(n_B(k_3)-p_3\frac{\partial n_B(k_3)}{\partial k_3}+n_F(k_3)\right) \left(\frac{-2p_3}{p_0^2-p_3^2}\right)\Bigg].
\eea
The fermion part of $I_1$ can be written as
\bea
-\int_{-\infty}^{\infty} dk_3 \frac{n_F(k_3)}{4k_3}\left(\frac{1}{k_3}+\frac{2p_3}{p_0^2-p_3^2}\right)
&\approx&  -\int_{0}^{\infty} dk_3\,\,\frac{1}{4} \left( \frac{2n_F}{k_3}\frac{2p_3}{p_0^2-p_3^2}\right)\nn\\
&=& \frac{p_3}{p_0^2-p_3^2}\left[\frac{1}{4\eps}+ \frac{\gamma_E}{2}+\frac{1}{2}\ln{\frac{2}{\pi}}\right]+\mathcal{O}[\eps]\label{fermion_I1}.
\eea
The bosonic part of $I_1$ is given as
\bea
&&\int_{-\infty}^{\infty} dk_3\,\frac{1}{4k_3}\left(n_B(k_3)-p_3 \frac{dn_B}{dk_3}\right)\left(\frac{1}{k_3}-\frac{2p_3}{p_0^2-p_3^2}\right)\nn\\
&\approx& \frac{1}{4}\left[-2\int_{0}^{\infty} dk_3\,\frac{n_B(k_3)}{k_3}\frac{2p_3}{p_0^2-p_3^2}-2\int_{0}^{\infty} dk_3\,\frac{p_3}{k_3^2}\frac{dn_B}{dk_3}\right]\nn\\
&=& \frac{1}{4}\left[-2\int_{0}^{\infty} dk_3\,\frac{n_B(k_3)}{k_3}\frac{2p_3}{p_0^2-p_3^2}-\frac{2p_3}{T}\frac{\partial}{\partial \beta}\int_{0}^{\infty} dk_3\,\frac{n_B(k_3)}{k_3^3}\right]\nn\\
&=&  \frac{p_3}{p_0^2-p_3^2}\left(-\frac{1}{4\eps}- \frac{\gamma_E}{2}+\frac{1}{2}\ln{2\pi}\right)-\frac{p_3\zeta'[-2]}{2T^2}+\mathcal{O}[\eps] . \label{bosonic_I1}
\eea
After combining \eqref{fermion_I1} and \eqref{bosonic_I1}, $I_1$ can be written as
\bea
I_1&=&\frac{p_3}{p_0^2-p_3^2}\ln{2}-\frac{p_3\zeta'[-2]}{2T^2},
\eea
\bea
\int_{-\infty}^{\infty} dk_3 \hspace{.2cm} k_3\, T_3 &=&- \frac{1}{2} \frac{\partial}{\partial p_3}\int  dk_3\,T_1\nn\\
&\approx&- \frac{1}{2} \frac{\partial}{\partial p_3}\Bigg[-\frac{1}{2}\int_{0}^{\infty} dk_3\,\frac{n_F(k_3)}{k_3^3}+\frac{1}{2}\int_{0}^{\infty} dk_3\,\,\frac{n_B(k_3)}{k_3^3}\nn\\
&&+\frac{p_3}{T(p_0^2-p_3^2)}\frac{\partial}{\partial \beta}\int_{0}^{\infty} dk_3\,\frac{n_B(k_3)}{k_3^3}\Bigg]\nn\\
&=&-\frac{\zeta[-2]}{T^2}\frac{p_0^2p_3}{(p_0^2-p_3^2)^2}.
\eea

So, 
\bea
b=-c=\frac{g^2 C_F (q_f B)}{4 \pi^2}\bigg[ \frac{p_3}{p_0^2-p_3^2} \ln 2  -\frac{p_3}{2 T^2} \zeta^\prime(-2)  - q_f B 
\frac{\zeta^\prime(-2)}{T^2} \frac{p_0^2 p_3}{(p_0^2-p_3^2)^2}\bigg].
\eea
\section{One-loop sum-integrals for quark free energy}
\label{quark_free_energy}
We write the form factors $ a$ and $b$ as
\bea
a&=&c_1\bigg[ \frac{p_0}{p_0^2-p_3^2} c_2 -d_1 \frac{p_0(p_0^2+p_3^2)}{(p_0^2-p_3^2)^2}\bigg]\nn\\
&=& c_1\bigg[ \frac{p_0}{P^2} c_2-d_1\(\frac{p_0}{P^2}+\frac{2p_0p_3^2}{P^4}\)\bigg],\nn\\
b&=& -c_1\bigg[ \frac{p_3}{p_0^2-p_3^2} c_2  -p_3 d_2 - 2d_1 \frac{p_0^2 p_3}{(p_0^2-p_3^2)^2}\bigg]\nn\\
&=& -c_1\bigg[ \frac{p_3}{P^2} c_2 -p_3 d_2 -2 d_1 \(\frac{p_3}{P^2}+\frac{p_3^3}{P^4}\)\bigg],
\eea
where
\bea
c_1&=&-\frac{g^2 C_F(q_fB)}{4 \pi^2}\,\,,\,\,c_2=\ln{2}\,\,,\,\,d_1=q_fB\frac{\zeta^\prime(-2)}{2 T^2}\,\,,\,\,
d_2=\frac{1}{2 T^2} \zeta^\prime(-2).
\eea
Now one can write the following frequency sum as
‌\bea
\sumintf_{\{p_0\}} ~\frac{ap_0}{P_{\sp}^2}&=& c_1 \sumintf_{\{p_0\}} \bigg[\(c_2-d_1\)\frac{1}{P_{\sp}^2}+\(c_2-d_1\)\frac{p_3^2}{P_{\sp}^4}-2d_1\(\frac{p_3^2}{P_{\sp}^4}+\frac{p_3^4}{P_{\sp}^6}\)\bigg],\\
\sumintf_{\{p_0\}} ~\frac{bp_3}{P_{\sp}^2}&=&-c_1 \sumintf_{\{p_0\}} \bigg[  \(c_2-2d_1\)\frac{p_3^2}{P_{\sp}^4}-d_2\frac{p_3^2}{P_{\sp}^2} -2d_1 \frac{p_3^4}{P_{\sp}^6}\bigg],\\
\sumintf_{\{p_0\}} \frac{a^2}{P_{\sp}^2}&=&c_1^2 \sumintf_{\{p_0\}} \bigg[\frac{p_0^2}{P_{\sp}^6}(\ln 2)^2 +d_1^2 \,\frac{p_0^2(p_0^2+p_3^2)^2}{P^{10}}-2d_1 \frac{p_0^2(p_0^2+p_3^2)}{P_{\sp}^8}\ln 2\bigg]\nn\\
&=& c_1^2\sumintf_{\{p_0\}} \bigg[(c_2-d_1)^2\frac{1}{P_{\sp}^4}+\(c_2^2+5d_1^2-6c_2 d_1\)\frac{p_3^2}{P_{\sp}^6}+4d_1\(2d_1-c_2\)\nn\\
&&~~~~~~~~~\times\frac{p_3^4}{P_{\sp}^8}+4d_1^2\frac{p_3^6}{P_{\sp}^{10}}\bigg] ,\\
\sumintf_{\{p_0\}} \frac{b^2}{P_{\sp}^2}&=&c_1^2\sumintf_{\{p_0\}}\bigg[\(c_2-2d_1\)^2\frac{p_3^2}{P_{\sp}^6}+4d_1^2\frac{p_3^6}{P_{\sp}^{10}}+4d_1\(2d_1-c_2\)\frac{p_3^4}{P_{\sp}^8}\nn\\
&&~~~~~~~+2d_2\(2d_1-c_2\)\frac{p_3^2}{P_{\sp}^4}+4d_1d_2\frac{p_3^4}{P_{\sp}^6}\bigg],\\
\sumintf_{\{p_0\}} ~\frac{a^2p_3^2}{P_{\sp}^4}&=&c_1^2\sumintf_{\{p_0\}} \bigg[(c_2-d_1)^2\frac{p_3^2}{P_{\sp}^6}+\(c_2^2+5d_1^2-6c_2 d_1\)\frac{p_3^4}{P_{\sp}^8}+4d_1\(2d_1-c_2\)\nn\\
&&~~~~~~~~\times\frac{p_3^6}{P_{\sp}^{10}}+4d_1^2\frac{p_3^8}{P_{\sp}^{12}}\bigg],\\
\sumintf_{\{p_0\}} ~\frac{b^2p_3^2}{P_{\sp}^4}&=&c_1^2\sumintf_{\{p_0\}}\bigg[\(c_2-2d_1\)^2\frac{p_3^4}{P_{\sp}^8}+4d_1^2\frac{p_3^8}{P_{\sp}^{12}}+4d_1\(2d_1-c_2\)\frac{p_3^6}{P_{\sp}^{10}}\nn\\
&&~~~~~~~+2d_2\(2d_1-c_2\)\frac{p_3^4}{P_{\sp}^6}+4d_1d_2\frac{p_3^6}{P_{\sp}^8}\bigg],\\
\sumintf_{\{p_0\}} ~\frac{abp_0p_3}{P_{\sp}^4}&=&c_1^2\sumintf_{\{p_0\}}\bigg[\(c_2-d_1\)\(c_2-2d_1\)\frac{p_3^2}{P_{\sp}^6}-d_2\(c_2-d_1\)\frac{p_3^2}{P_{\sp}^4}\nn\\
&&~~~~~~~~+\{\(c_2-d_1\)\(c_2-6d_1\)+2d_1^2\}\frac{p_3^4}{P_{\sp}^8}+d_2\(3d_1-c_2\)\frac{p_3^4}{P_{\sp}^6}\nn\\
&&~~~~~~~~~+2d_1\(5d_1-2c_2\)\frac{p_3^6}{P_{\sp}^{10}}+2d_1d_2\frac{p_3^6}{P_{\sp}^8}+4d_1^2\frac{p_3^8}{P_{\sp}^{12}}\bigg].
\eea
So $F'_q$ in Eq.~{\eqref{F_2_exp}} becomes
\bea
F'_q&=& -4 d_F \sum_f \frac{q_fB}{(2\pi)^2}\sumintf_{\{p_0\}}dp_3~\bigg[\frac{a p_{0}}{P_{\sp}^{2}}+\frac{b p_{3}}{P_{\sp}^{2}}
-\frac{a^{2}}{P_{\sp}^{2}}-\frac{b^{2}}{P_{\sp}^{2}}-\frac{2 a^{2} p_{3}^{2}}{P_{\sp}^{4}}-\frac{2 b^{2} p_{3}^{2}}{P_{\sp}^{4}}\nn\\
&&~~~~~~~~~~~~~~~~~~~~~~~~~~~~~~~~~-\frac{4 a b p_{0} p_{3}}{P_{\sp}^{4}}\bigg]\\
&=&-4 d_F \sum_f \frac{q_fB}{(2\pi)^2} \bigg[c_1(c_2-d_1)~ \sumintf_{\{p_0\}}~\frac{1}{P_{\sp}^{2}}-c_1\big(d_1 - 6 c_1 c_2 d_2 \nn\\
&&+ 8 c_1 d_1 d_2\big)\sumintf_{\{p_0\}}~\frac{p_3^2}{P_{\sp}^4}+8 c_1^2 d_2\(c_2 - 3 d_1\)\sumintf_{\{p_0\}}~\frac{p_3^4}{P_{\sp}^{6}}-16c_1^2d_1d_2 \sumintf_{\{p_0\}}~\frac{p_3^6}{P_{\sp}^{8}}\nn\\&&+c_1d_2\sumintf_{\{p_0\}}~\frac{p_3^2}{P_{\sp}^{2}}-c_1^2\(c_2-d_1\)^2\sumintf_{\{p_0\}}~\frac{1}{P_{\sp}^{4}}-c_1^2 \(8 c_2^2 - 26 c_2 d_1 + 19 d_1^2\)\nn\\
&&\sumintf_{\{p_0\}}~\frac{p_3^2}{P_{\sp}^{6}}-2 c_1^2 \(2 c_2 - 11 d_1\) \(2 c_2 - 3 d_1\)\sumintf_{\{p_0\}}~\frac{p_3^4}{P_{\sp}^{8}}-16 c_1^2 d_1 \(-2 c_2 + 5 d_1\)\nn\\&&\times\sumintf_{\{p_0\}}~\frac{p_3^6}{P_{\sp}^{10}}-32c_1^2d_1^2\sumintf_{\{p_0\}}~\frac{p_3^8}{P_{\sp}^{12}}\bigg]. \label{quark_fe}
\eea
Now we calculate the frequency sums as
\bea
\sum_{\{p_0\}} \frac{1}{P_{\sp}^4}&=&\frac{1}{2 p_3}\frac{\partial}{\partial p_3}\sum_{\{p_0\}} \frac{1}{P_{\sp}^2}\label{P4_sum_F},\nn\\
\sum_{\{p_0\}} \frac{1}{P_{\sp}^6}&=&\frac{1}{4 p_3}\frac{\partial}{\partial p_3}\sum_{\{p_0\}} \frac{1}{P_{\sp}^4},\nn\\
\sum_{\{p_0\}} \frac{1}{P_{\sp}^8}&=&\frac{1}{6 p_3}\frac{\partial}{\partial p_3}\sum_{\{p_0\}} \frac{1}{P_{\sp}^6},\nn\\
\sum_{\{p_0\}} \frac{1}{P_{\sp}^{10}}&=&\frac{1}{8 p_3}\frac{\partial}{\partial p_3}\sum_{\{p_0\}} \frac{1}{P_{\sp}^8}.
\eea
One can calculate
\bea
 \sum_{\{p_0\}} \frac{1}{P_{\sp}^2}&=& -\frac{1}{2 p_3}\bigg(1-2 \, n_F(p_3)\bigg)\label{P2_sum_F},
\eea
Thus using Eq.~\eqref{P2_sum_F} in Eq.~\eqref{P4_sum_F}, one can write
\bea
 \sum_{\{p_0\}} \frac{1}{P_{\sp}^4}\!&=&\! \frac{1}{2 p_3}\frac{\partial}{\partial p_3}\bigg( \sum_{\{p_0\}} \frac{1}{P_{\sp}^2}\bigg)
\!\approx\!\frac{1}{2 p_3} \frac{\partial}{\partial p_3} \bigg[\frac{n_F(p_3)}{p_3}\bigg]
\!\!=\!\! \frac{1}{2 p_3}\bigg[ \frac{ \beta}{p_3^2} \frac{\partial n_F(p_3)}{\partial \beta}-\frac{n_F(p_3)}{p_3^2}\bigg].
\eea
Now we perform the sum-integrals in Eq.~\eqref{quark_fe} as
\bea
 \sumintf_{\{p_0\}} \, \frac{1}{P_{\sp}^4}&=&  \(\frac{e^{\gamma_E}\Lambda^2}{4\pi }\)^{\epsilon}\int_{-\infty}^{\infty} d^{1-2\eps}p_3\left(  \beta \frac{\partial}{\partial \beta}-1\right)\frac{n_F(p_3)}{2p_3^3} \nn\\
&\approx& 2 \(\frac{e^{\gamma_E}\Lambda^2}{4\pi }\)^{\epsilon}\int_{0}^{\infty} d^{1-2\eps}p_3\left(  \beta \frac{\partial}{\partial \beta}-1\right)\frac{n_F(p_3)}{2p_3^3}\nn\\
&\approx& \(\frac{\Lambda}{4\pi T}\)^{2\eps}\[-\frac{7}{2} \frac{\zeta^{'} (-2)}{T^2}+O(\epsilon)\]  ,
\eea
\bea
 \sumintf_{\{p_0\}} \, \frac{p_3^2}{P_{\sp}^6}
&=&  \(\frac{e^{\gamma_E}\Lambda^2}{4\pi }\)^{\epsilon}{\int_{-\infty}^{\infty} d^{1-2\eps}p_3\left(\beta^2 \frac{\partial^2}{\partial \beta^2}-3\beta\frac{\partial}{\partial \beta}+3\right)\frac{n_F(p_3)}{8p_3^3}}\nn\\
&\approx&\(\frac{\Lambda}{4\pi T}\)^{2\eps}\[ {\frac{7}{8}\frac{ \zeta^{'}(-2)}{T^2}+O(\eps)}\],\\
 \sumintf_{\{p_0\}} \, \frac{p_3^4}{P_{\sp}^8}&=&  \(\frac{e^{\gamma_E}\Lambda^2}{4\pi }\)^{\epsilon}\int_{-\infty}^{\infty}  d^{1-2\eps}p_3 \left(\beta^3 \frac{\partial^{3}}{\partial \beta^{3}}-6\beta^2\frac{\partial^{2}}{\partial \beta^{2}}+15\beta\frac{\partial}{\partial \beta}-15\right)\frac{n_F(p_3)}{48 p_3^3}\nn\\
&\approx&\(\frac{\Lambda}{4\pi T}\)^{2\eps}\[-\frac{7}{16}\frac{\zeta^{'}(-2)}{T^2}+O(\eps)\],\\
 \sumintf_{\{p_0\}} \, \frac{p_3^6}{P_{\sp}^{10}}&=& \(\frac{e^{\gamma_E}\Lambda^2}{4\pi }\)^{\epsilon} \int_{-\infty}^{\infty} d^{1-2\eps}p_3 \bigg(\beta^4\frac{\partial^{4}}{\partial \beta^{4}}-10\beta^3\frac{\partial^{3}}{\partial \beta^{3}}+45\beta^2\frac{\partial^{2}}{\partial \beta^{2}}-105\beta\frac{\partial}{\partial \beta}\nn\\
 &&+105\bigg)\frac{n_F(p_3)}{384 p_3^3}\nn\\
&\approx&\(\frac{\Lambda}{4\pi T}\)^{2\eps}\[\frac{35}{256}\frac{\zeta^{'}(-2)}{T^2}+O(\eps)\],\\
 \sumintf_{\{p_0\}} \, \frac{p_3^8}{P_{\sp}^{12}}&=& \(\frac{e^{\gamma_E}\Lambda^2}{4\pi }\)^{\epsilon}\int_{-\infty}^{\infty} d^{1-2\eps}p_3 \bigg(\beta^5\frac{\partial^{5}}{\partial \beta^{5}}-15\beta^4\frac{\partial^{4}}{\partial \beta^{4}}+105\beta^3\frac{\partial^{3}}{\partial \beta^{3}}\nn\\
 &&-420\beta^2\frac{\partial^{2}}{\partial \beta^{2}}+945\beta\frac{\partial}{\partial \beta}-945\bigg)\frac{n_F(p_3)}{3840 p_3^3}\nn\\
&\approx&\(\frac{\Lambda}{4\pi T}\)^{2\eps}\left[-\frac{49}{256}\frac{\zeta'(-2)}{T^2}+O(\eps)\right].\\
\eea
Similarly one can calculate
\bea
 \sumintf_{\{p_0\}} \, \frac{1}{P_{\sp}^{2}}&=&\(\frac{e^{\gamma_E}\Lambda^2}{4\pi }\)^{\epsilon}\int_{-\infty}^{\infty} d^{1-2\eps}p_3\, \frac{n_F(p_3)}{p_3}\nn\\
&\approx&\(\frac{\Lambda}{4\pi T}\)^{2\eps}\left[-\frac{1}{2\eps}-\frac{1}{2}\(3\gamma_E+4\ln{2}-\ln{\pi}\)+O(\eps)\right],\\
 \sumintf_{\{p_0\}} \, \frac{p_3^2}{P_{\sp}^{4}}&=&\(\frac{e^{\gamma_E}\Lambda^2}{4\pi }\)^{\epsilon}\int_{-\infty}^{\infty} d^{1-2\eps}p_3\, \(\beta\frac{\partial}{\partial \beta}-1\)\frac{n_F(p_3)}{2p_3}\nn\\
&\approx&\(\frac{\Lambda}{4\pi T}\)^{2\eps}\left[\frac{1}{4\eps}+\frac{1}{4}\(-2+3\gamma_E+4\ln{2}-\ln{\pi}\)+O(\eps)\right],\\
 \sumintf_{\{p_0\}} \, \frac{p_3^4}{P_{\sp}^{6}}&=&\(\frac{e^{\gamma_E}\Lambda^2}{4\pi }\)^{\epsilon}\int_{-\infty}^{\infty} d^{1-2\eps}p_3\, \(\beta^2\frac{\partial^2}{\partial \beta^2}-3\beta \frac{\partial}{\partial \beta}+3\)\frac{n_F(p_3)}{8p_3}\nn\\
&\approx&\(\frac{\Lambda}{4\pi T}\)^{2\eps}\left[-\frac{3}{16\eps}-\frac{3}{16}\(-\frac{8}{3}+3\gamma_E+4\ln{2}-\ln{\pi}\)+O(\eps)\right],\\
 \sumintf_{\{p_0\}} \, \frac{p_3^6}{P_{\sp}^{8}}&=&\(\frac{e^{\gamma_E}\Lambda^2}{4\pi }\)^{\epsilon}\int_{-\infty}^{\infty} d^{1-2\eps}p_3\, \(\beta^3\frac{\partial^3}{\partial \beta^3}-6\beta^2\frac{\partial^2}{\partial \beta^2}+15\beta\frac{\partial}{\partial \beta}-15\)\frac{n_F}{48p_3}\nn\\
&\approx&\(\frac{\Lambda}{4\pi T}\)^{2\eps}\left[\frac{5}{32\eps}+\frac{5}{32}\(-\frac{46}{15}+3\gamma_E+4\ln{2}-\ln{\pi}\)+O(\eps)\right],\\
 \sumintf_{\{p_0\}} \, \frac{p_3^2}{P_{\sp}^{2}}&=& \(\frac{\Lambda}{4\pi T}\)^{2\eps}\left[\frac{\pi^{2}T^{2}}{6}+O(\eps)\right].
\eea
Using the above sum-integrals in Eq.~(\ref{quark_fe}) $F'_q$ up to $\mathcal{O}(g^4)$ can be written as,
\bea
F'_q
&=& -4d_F \sum_f \frac{\(q_f B\)^2}{(2\pi)^2}\frac{g^2 C_F}{4 \pi^2}\(\frac{\Lambda}{4\pi T}\)^{2\eps}\Bigg[\frac{1}{8\eps}  \(4\ln{2} - q_fB\frac{\zeta^{'} (-2)}{T^2} \)+\frac{1}{24576}\nn\\
&&\times\bigg\{12288 \ln{2} (3 \gamma_E + 4\ln{2}-\ln{\pi})+\frac{256\zeta[3]}{\pi^4T^2}\Big(2 \pi^4 T^2 -3 g^2 C_F (q_fB) \ln{2}\nn\\
&&+ 
 3 \pi^2 (q_fB)(2 + 3 \gamma_E + 4\ln{2}- \ln{\pi})\Big)-\frac{8g^2C_F}{\pi^6T^4}(q_FB)^2 \zeta[3]^2(4 + 105 \ln{2})\nn\\
 && +\frac{7245g^2C_F}{\pi^8T^6}(q_FB)^3 \zeta[3]^3\bigg\}\Bigg].
\eea
\section{HTL One-loop sum-integrals for gluon free energy}
\label{HAS}
\bea
\sumintb_{P} \frac{1}{P^2}&=&-\frac{T^2}{12}\left(\frac{\Lambda}{4\pi T}\right)^{2\eps}\Bigg[1+2\eps\left(1+\frac{\zeta'(-1)}{\zeta(-1)}\right)\Bigg]+\mathcal{O}[\eps]^2,\label{gluon_1}\\
\sumintb_P \frac{1}{p^2P^2} &=& -\frac{2}{(4\pi)^2}\left(\frac{\Lambda}{4\pi T}\right)^{2\eps}\Bigg[\frac{1}{\eps}+2\gamma_E+2
+\eps\left(4+4\gamma_E+\frac{\pi^2}{4}-4\gamma_1\right)\Bigg]\nn\\
&&+\mathcal{O}[\eps]^2,\label{gluon_2}\\
\sumintb_P \frac{1}{P^4} &=&  \frac{1}{(4\pi)^2}\left(\frac{\Lambda}{4\pi T}\right)^{2\eps}\left[\frac{1}{\eps}
+2\gamma_E+\eps\left(\frac{\pi^2}{4}-4\gamma_1\right)\right]+\mathcal{O}[\eps]^2,\\
\sumintb_{P}\frac{\mathcal{T}_p}{p^2P^2}&=&\sumintb_{P}\left\langle \frac{1-c^{2\eps+1}}{1-c^2}\right\rangle_c \frac{1}{p^2P^2}\nn\\
&=&-\frac{2}{(4\pi)^2}\left(\frac{\Lambda}{4\pi T}\right)^{2\eps} \(\ln 2+\(\frac{\pi^2}{6}-(2-\ln 2)\ln 2\)\eps\)\Bigg[\frac{1}{\eps}+2\gamma_E+2\nn\\
&&+\eps\bigg(4+4\gamma_E+\frac{\pi^2}{4}-4\gamma_1\bigg)\Bigg]\nn\\
&=&-\frac{2}{(4\pi)^2}\left(\frac{\Lambda}{4\pi T}\right)^{2\eps}\Bigg[\frac{1}{\epsilon }\ln 2+\frac{\pi ^2}{6}+\(\ln 2\)^2+2\gamma_E  \ln 2\Bigg],\label{gluon_3}\\
\sumintb_{P}\frac{\mathcal{T}_p}{p^4}&=&\sumintb_{P} \left\langle c^{2\eps+1}\right\rangle_c \frac{1}{p^2P^2}\nn\\
&=& -\frac{2}{(4\pi)^2}\left(\frac{\Lambda}{4\pi T}\right)^{2\eps}\left\{\frac{1}{2}+\(-1 + \ln 2\)\eps\right\} \Bigg[\frac{1}{\eps}+2\gamma_E+2\nn\\
&&+\eps\bigg(4+4\gamma_E+\frac{\pi^2}{4}-4\gamma_1\bigg)\Bigg]\nn\\
&=&-\frac{2}{(4\pi)^2}\left(\frac{\Lambda}{4\pi T}\right)^{2\eps}\Bigg[\frac{1}{2 \eps} + \(\gamma_E + \ln 2\)\Bigg],\label{gluon_4}\\
\sumintb_{P}\frac{\mathcal{T}_p^2}{p^4}&=&\sumintb_{P}\left\langle \frac{c_1^{2\eps+3}-c_2^{2\eps+3}}{c_1^2-c_2^2}\right\rangle_c \frac{1}{p^2P^2}\nn\\
&=&-\frac{2}{(4\pi)^2}\left(\frac{\Lambda}{4\pi T}\right)^{2\eps}\(\frac{1}{3}(1+2\ln2) + \frac{2}{9}\(-5+\ln 2\(5+3\ln 2\)\)\eps\)\nn\\
&&\times\Bigg[\frac{1}{\eps}+2\gamma_E+2
+\eps\left(4+4\gamma_E+\frac{\pi^2}{4}-4\gamma_1\right)\Bigg]\nn\\
&=&-\frac{2}{(4\pi)^2}\left(\frac{\Lambda}{4\pi T}\right)^{2\eps}\Bigg[\frac{1}{3\eps}\(1+2\ln 2\)+\frac{2}{9} \bigg(-2+3 \(\ln 2\)^2\nn\\
&&+\gamma_E  (3+6\ln 2)+11\ln 2\bigg)\Bigg]\label{gluon_5}.
\eea

\bea
&&\sumintb_{P}e^{{-p_\perp^2}/{2q_fB}}\frac{1}{p^2P^2}\frac{p_3^2}{\(p_0^2-p_3^2\)}\nn\\
&=&\sumintb_{P}e^{{-p_\perp^2}/{2q_fB}}\frac{p_3^2}{p^2\(p^2-p_3^2\)}\left\{\frac{1}{P^2}-\frac{1}{p_0^2-p_3^2}\right\}\nn\\
&=&\sumintb_{P}e^{{-p_\perp^2}/{2q_fB}} \frac{p_3^2}{p^2\(p^2-p_3^2\)P^2}-\sumintb_{P}e^{{-p_\perp^2}/{2q_fB}} \frac{p_3^2}{p^2\(p^2-p_3^2\)\(p_0^2-p_3^2\)}\nn\\
&=&\(\frac{e^{\gamma_E}\Lambda^2}{4\pi }\)^{\epsilon}\int\frac{d^{3-2\epsilon}p}{(2\pi)^{3-2\epsilon}} \Bigg[-e^{{-p_\perp^2}/{2q_fB}} \frac{p_3^2}{p^3\(p^2-p_3^2\)}
n_B(p)+e^{{-p_\perp^2}/{2q_fB}}\nn\\
&&~~~~~~~~~~~~~~~\times\frac{p_3}{p^2\(p^2-p_3^2\)}n_B(p_3)\Bigg]\nn\\
&\approx&\(\frac{\Lambda}{4\pi T}\)^{2 \eps}\frac{1}{(4\pi)^29q_fB}\bigg[\frac{1}{\eps}\bigg(18 q_fB - 3 \pi^2 T^2 - 18 q_fB \ln 2\bigg)+\bigg\{-3 qfB \nn\\
&&\bigg(-12 + \pi^2 + 6 (\ln 2)^2 + 
    6 \gamma_E (-2 + 2\ln 2)\bigg) - \pi^2 T^2 \bigg(-8 + 6\ln 2 \nn\\
    &&+ 
    6\(1+\frac{\zeta'[-1]}{\zeta[-1]}\) \bigg)\bigg\}\bigg]+O\bigg[\frac{1}{(q_f B)^2}\bigg].\label{gluon_6}
\eea

\bea
&&\sumintb_{P}e^{{-p_\perp^2}/{2q_{f_1}B}}e^{{-p_\perp^2}/{2q_{f_2}B}}\frac{p_3^4}{p^4\(p_0^2-p_3^2\)^2}\nn\\
&=&-\(\frac{e^{\gamma_E}\Lambda^2}{4\pi }\)^{\epsilon}\int\frac{d^{3-2\epsilon}p}{(2\pi)^{3-2\epsilon}} e^{{-p_\perp^2}/{2q_{f_1}B}}e^{{-p_\perp^2}/{2q_{f_2}B}}\frac{p_3}{2p^4}\(\beta \frac{\partial}{\partial \beta}-1\)n_B(p_3)\nn\\
&\approx&-\(\frac{\Lambda}{4\pi T}\)^{2 \eps}\frac{1}{(4 \pi)^2}\bigg[\frac{T^4}{36(q_{f_1}B)^2(q_{f_2}B)^2\eps}\bigg\{-\frac{18 {(q_{f_1}B)}^2{(q_{f_2}B)}^2}{T^4}+\frac{18 \pi ^2 {(q_{f_1}B)}{q_{f_2}B}}{T^2}\nn\\
&&\({q_{f_1}B}+{q_{f_2}B}\)+6 \pi ^4 \({(q_{f_1}B)}^2+{(q_{f_2}B)}^2\)+12 \pi ^4 {q_{f_1}B} {q_{f_2}B}\bigg\}+\frac{1}{36(q_{f_1}B)^2(q_{f_2}B)^2}\nn\\
&&\bigg\{-18 \bigg( {(q_{f_1}B)}^2 \left( {(q_{f_2}B)}^2 \ln 4+12  {q_{f_2}B} T^2 \zeta '[2]+60 T^4 \zeta '[4]\right)+12 {q_{f_1}B} {q_{f_2}B} T^2\nn\\
&&
   \left({q_{f_2}B} \zeta '[2]+10 T^2 \zeta '[4]\right)+60 {(q_{f_2}B)}^2 T^4 \zeta '[4]\bigg)+12 \gamma_E  \big({(q_{f_1}B)}^2 \Big(-3
   {(q_{f_2}B)}^2\nn\\
   &&+3 \pi ^2 {q_{f_2}B} T^2+\pi ^4 T^4\Big)+\pi ^2 {q_{f_1}B} {q_{f_2}B} T^2 \left(3 {q_{f_2}B}+2 \pi ^2 T^2\right)+\pi
   ^4 {(q_{f_2}B)}^2 T^4\big)\nn\\
   &&+\pi ^4 T^4 (12 \ln (4 \pi )-25) ({q_{f_1}B}+{q_{f_2}B})^2+18 \pi ^2 {q_{f_1}B} {q_{f_2}B} T^2 (-3+2 \ln (4\pi )) \nn\\
   &&({q_{f_1}B}+{q_{f_2}B})\bigg\}\bigg]+O\bigg[\frac{1}{(q_f B)^3}\bigg].\label{gluon_7}
\eea
\bea
&&\sumintb_{P}e^{{-p_\perp^2}/{2q_fB}} \frac{p_3^2}{p^4\(p_0^2-p_3^2\)}
= -\(\frac{e^{\gamma_E}\Lambda^2}{4\pi }\)^{\epsilon}\int\frac{d^{3-2\epsilon}p}{(2\pi)^{3-2\epsilon}}e^{{-p_\perp^2}/{2q_fB}} \frac{p_3}{p^4}n_B(p_3)\nn\\
&\approx&-\(\frac{\Lambda}{4\pi T}\)^{2 \eps}\frac{1}{(4\pi)^2}\bigg[\frac{1}{\eps} \left(1 - \frac{\pi^2 T^2}{3 q_fB} \right)+\frac{1}{3}\bigg\{-\frac{\pi ^2 T^2 }{q_fB}\(2\(1+\frac{\zeta'[-1]}{\zeta[-1]}\)-3+2\ln 2\) \nn\\
&&+6 \gamma_E+6\ln 2\bigg\}\bigg]+O\bigg[\frac{1}{(q_f B)^2}\bigg].\label{gluon_8}
\eea

\bea
&&\sumintb_{P}e^{{-p_\perp^2}/{2q_fB}}\frac{p_3^2\mathcal{T}_p}{p^4\(p_0^2-p_3^2\)}\nn\\
&=& \(\frac{e^{\gamma_E}\Lambda^2}{4\pi }\)^{\epsilon}\int\frac{d^{3-2\epsilon}p}{(2\pi)^{3-2\epsilon}}e^{{-p_\perp^2}/{2q_fB}}\Bigg\langle -\frac{p_3n_B(p_3)}{p^4}-\frac{p_3 c^2 n_B(p_3)}{p^2\(p_3^2-p^2c^2\)}
+\frac{p_3^2 c  n_B(pc)}{p^3\(p_3^2-p^2c^2\)}\Bigg\rangle_c \nn\\
&\approx&\(\frac{\Lambda}{4\pi T}\)^{2 \eps}\frac{1}{(4\pi)^2}\bigg[\frac{1}{\eps} \bigg(\frac{1}{3} - \frac{2 \pi^2 T^2}{9 q_fB}\bigg)+\frac{1}{27q_fB}\bigg\{\pi ^2 T^2 \bigg(-12 \(1+\frac{\zeta'[-1]}{\zeta[-1]}\)\nn\\
&&+17+12\ln 2\bigg)-3 q_fB \left(-8+12 (\ln 2)^2+18 \ln 2+6 \gamma_E  (4\ln 2-1)+8\ln 2\right)\bigg\}\bigg]\nn\\
&&+O\bigg[\frac{1}{(q_f B)^2}\bigg].
\label{gluon_9}
\eea
Using Eqs.~(\ref{gluon_1})-(\ref{gluon_9}) in Eq.~(\ref{fe_sfa}) one can get an expression for gluon free energy in a strongly magnetized medium computed up to $\mathcal O[g^4]$ as
\bea
 F_g
&=& d_A\frac{1}{(4\pi)^2}\Bigg[\frac{1}{\eps}\Bigg\{-\frac{1}{8}\(\frac{C_A g^2T^2}{3}\)^2+\sum_{f_1,f_2} \frac{g^4T^4}{192(q_{f_1}B)(q_{f_2}B)}\bigg( {(q_{f_1}B)}^2+{(q_{f_2}B)}^2\bigg)\nn\\
&&+\frac{N_f^2g^4T^4}{96}+ \frac{C_A N_f g^4T^4}{36}-\sum_{f_1,f_2} \frac{g^4(q_{f_1}B)(q_{f_2}B)}{64\pi^4}+\sum_{f_1,f_2} \frac{g^4T^2}{64\pi^2}\Big({q_{f_1}B}+{q_{f_2}B}\Big)\nn\\
&&-\sum_f \frac{1}{4\pi^2}\frac{C_A g^4T^2q_fB}{6}\(1+\ln 2\)\Bigg\}-\frac{16 \pi^4 T^4}{45}+\frac{2C_A g^2\pi^2T^4}{9}+\(\frac{C_A g^2T^2}{3}\)^2\nn\\
&&\times \frac{1}{12}\Big(8 - 3 \gamma_E - \pi^2 + 7\ln 2 - 3 \ln \hat \Lambda\Big)+\frac{1}{2}\sum_{f_1,f_2} \(\frac{g^2}{4\pi^2}\)^2\Bigg\{\frac{1}{2} \pi ^2 T^2 \bigg(2\(1+\frac{\zeta'[-1]}{\zeta[-1]}\)\nn\\
&&-3+2\ln 2\bigg) (q_{f_1}B+q_{f_2}B)+(2\ln 2-2 \ln \hat \Lambda ) \bigg(-\frac{\pi ^4 T^4}{6 (q_{f_1}B) (q_{f_2}B)}
   \Big({(q_{f_1}B)}^2\nn\\
   &&+{(q_{f_2}B)}^2\Big)-\frac{1}{2} \pi ^2 T^2 (q_{f_1}B+q_{f_2}B)+\frac{(q_{f_1}B)
   (q_{f_2}B)}{2}-\frac{\pi ^4 T^4}{3}\bigg)-\frac{30 T^4 \zeta '[4]}{(q_{f_1}B)
   (q_{f_2}B)} \nn\\
   &&\times\big((q_{f_1}B)^2+(q_{f_2}B)^2\big)+\frac{\pi ^4 T^4 }{36 q_{f_1}B
  q_{f_2}B}(-25+12 \gamma_E +12 \ln (4 \pi )) \Big((q_{f_1}B)^2\nn\\
  &&+(q_{f_2}B)^2\Big)-q_{f_1}B q_{f_2}B (\gamma_E +\ln 2)-60 T^4 \zeta '[4]+\frac{1}{18} \pi ^4 T^4 (-25+12 \gamma_E \nn\\
  &&+12 \ln (4 \pi ))\Bigg\}-\sum_f \frac{g^2q_fB}{4\pi^2}\frac{C_A g^2T^2}{12}\Bigg\{\frac{T^2}{3q_fB}\bigg(12 \pi^2 - 8 \pi^2 \(1+\frac{\zeta'[-1]}{\zeta[-1]}\)\nn\\
  &&- 8 \pi^2 \ln\frac{\hat \Lambda}{2}\bigg)+\frac{1}{3} \bigg(4 (3+3\ln 2) \ln \frac{\hat \Lambda }{2}+\pi ^2-4-6 (\ln 2)^2-6 \gamma_E  (2\ln 2-2)\nn\\
  &&-8 \ln
   2\bigg)\Bigg\}\Bigg].
\eea

\end{document}